\documentclass[preprint,review,12pt]{elsarticle}

\usepackage{graphicx}
\usepackage{amsmath}
\usepackage{amssymb}
\usepackage{url}
\usepackage{wrapfig}
\usepackage[bookmarks=false,linkcolor=red]{hyperref}
\hypersetup{colorlinks=true,breaklinks,urlcolor=[rgb]{0,0,1},citecolor=[rgb]{0,0.6,0}}
\usepackage{color}
\usepackage{float}
\usepackage{geometry}
\begin{document}

\begin{frontmatter}


\title{A Physically Constrained Inversion for Super-resolved Passive Microwave Retrieval of Soil Moisture and Vegetation Water Content in L-band}

\author[1]{Ardeshir Ebtehaj\corref{cor1}}
\ead{ebtehaj@umn.edu}
\author[2]{Rafael L. Bras}

\cortext[cor1]{Corresponding author}

\address[1]{Saint Anthony Falls Laboratory, Department of Civil Environmental and Geo- Engineering, University of Minnesota, Minneapolis, Minnesota, 55414, United States.}

\address[2]{School of Civil and Environmental Engineering and School of Earth and Atmospheric Sciences, Georgia Institute of Technology, Atlanta , Georgia, 30332, United States.}
\vspace{-2mm}

\begin{abstract}

{\small Remote sensing of soil moisture and vegetation
water content from space often requires underdetermined inversion of a zeroth-order
approximation of the forward radiative transfer equation in L-band, known
as the $\tau$-$\omega$ model. This paper shows that the least-squares
inversion of the model is not strictly convex and the widely
used unconstrained damped least-squares (DLS) can lead to biased retrievals, chiefly due to the existing preferential solution spaces that are characterized
by the eigenspace of the model's Hessian. In particular,
the numerical experiments show that for sparse (dense) vegetation
with a shallow (deep) optical depth, the DLS tends to overestimate
(underestimate) the soil moisture and vegetation water content for
a dry (wet) soil. A new Constrained Multi-Channel Algorithm (CMCA) is proposed that bounds the retrievals with a priori information about the soil type and vegetation density and can account for slowly varying dynamics of the vegetation water content over croplands through a temporal smoothing-norm regularization in the derivative domain. It is demonstrated that depending on the resolution of the constraints, the algorithm can lead to super-resolved soil moisture retrievals beyond the spatial resolution of radiometric observations. Multiple Monte Carlo retrieval experiments are conducted and the results are validated against ground-based gauge observations.} 
\end{abstract}
\begin{keyword}
Microwaves Remote sensing, Soil Moisture Retrievals, Inverse problems, Regularization 
\end{keyword}

\end{frontmatter}


\section{Introduction}

Soil moisture is only less than 5 percent of the Earth's freshwater
reservoirs \cite{Gleick1998} but plays a key role in regulating
the water mass transport and energy exchange in the soil-plant-atmosphere
continuum.  The Earth's vegetation, and thus the global food
security, depends on the soil moisture climatology \cite{Liu2015}. Trends in intensity, frequency and duration of the global precipitation \cite{Trenberth_2011}, resulting from a changing climate, makes monitoring of soil moisture a key factor to extend forecast skill of land-atmosphere models \cite{Lin2016,Lin2017};
improve drought modeling and management \cite{Velpuri_2016,Shaw_2017};
and further unravel processes that regulate evapotranspiration
\cite{Katul_2012,Schlesinger2014}, as well as carbon \cite{Falloon2011}
and nitrogen cycles \cite{Pastor1986}. 

Emission of surface soil in microwave frequencies between 1 to 5 GHz
is sensitive to soil moisture content. The upwelling electromagnetic waves penetrate well through the overlaying canopy with moderate water content (i.e., $<$~5 kg\,m\textsuperscript{-2})
and reach to the top of the atmosphere with negligible interactions
with the atmospheric constituents \cite{Njoku1996a}. The L-band (1-2 GHz) radiometry has been
central to soil moisture satellite missions \citep{Vine_2007,Kerr2012} including the NASA's Soil Moisture Active Passive (SMAP, \cite{Entekhabi2010}) satellite. 

In L-band, the surface soil emission and its interaction with vegetation
biomass can be represented well through a zeroth-order nonlinear approximation
of the radiative transfer equation \citep[and references there in]{Njoku1977,Mo1982,Ulaby1982,Njoku1996a}---known as the $\tau$-$\omega$ model. This forward model relates the surface temperature to the observed brightness temperatures at the top of the atmosphere as a function of rough surface soil reflectivity, vegetation transmissivity and single scattering albedo. Thus, inversion of this
model could lead to the retrievals of not only the soil moisture but
also the vegetation water content (VWC) \citep[e.g.,][]{Jackson1991,Entekhabi1994,Njoku1999,Jackson2004}. Both of the soil reflectivity and vegetation transmissivity are polarization dependent; however, evidence suggests that the polarization dependence of the former is more pronounced than the latter in L-band. The single scattering albedo is the least dynamic parameter of the model and is often considered polarization
independent and seasonally invariant \citep{Griend_1993,Karam1992,Griend_96}. 

Simultaneous retrieval of the soil moisture and the vegetation transmissivity from single L-band radiometry could lead to an ill-posed nonlinear inverse problem. To make the problem well-posed, a priori information can be provided in different ways. In particular, there are two major classes of inversion algorithms---namely
single channel (SCA) and dual channel (DCA) algorithms. The SCA only uses horizontal or vertical polarization of the observed
brightness temperatures \citep{Jackson1993}. To make the inversion well-posed, prior information
is supplied through ancillary data. In particular, it is assumed that
the single scattering albedo is a known constant over different land
covers and the vegetation transmissivity can be estimated off-line
from climatology of the normalized difference vegetation index (NDVI).

The DCA  uses both polarization channels to compute the unknown parameters simultaneously, through a nonlinear least-squares (LS) inversion
\citep{Njoku1999}. Due to the strong dependence of the observed brightness
temperatures at different polarization channels, this inversion is often ill-conditioned, especially for the SMAP single band radiometer. To make the inversion possible,
the DCA often relies on the Levenberg-Marquardt (LM) or the damped
least-squares (DLS) optimization algorithm \citep{Levenberg1944a,Marquardt1963}.
This iterative algorithm can find a minimum solution of an ill-posed
nonlinear LS problem using an adaptive Tikhonov regularization \citep{Tikhonov1978}.  

Other advanced DCA inversion approaches have also been proposed that
rely on the DLS. \citet{Piles_2010} recast
the retrieval as an overdetermined inverse problem by proposing to
use the first guess of the free parameters as a-priori knowledge, assuming
that the uncertainties follow a Gaussian distribution.
More recently, \citet{Koning_2016} adopted a time series approach
that retrieves the real component of the soil dielectric constant
within a window of time---over which, it is assumed that
the vegetation optical depth remains constant. A review of existing algorithms can be found in the SMAP handbook \citep{Enrekhabi2014} and also in \citep{WIGNERON2017}.  

The contribution of this paper can be summarized as follows. First, we show that the LS inversion of the $\tau$-$\omega$ model is not strictly convex, which leads to a preferential solution space and biased DCA retrievals. Second, we propose to use a Tikhonov regularization to add box constraints to an LS inversion of the model that will lead to the retrievals that are physically consistent with the soil moisture capacity and climatology of vegetation phenology. Third, we extend the existing time-series retrieval algorithms \citep{Koning_2016} to account for the slow-varying changes in the vegetation water content, through a smoothing-norm regularization in a derivative domain. Fourth, we provide initial results that the algorithm can lead to higher resolution retrievals of soil moisture and vegetation water content than the native resolution of the SMAP radiometer, depending on the resolution of the constraints.    

Section \ref{sec:2} presents introductory materials
about the forward $\tau$-$\omega$ model and provides conceptual
details about its LS inversion. This section sets the stage for discussing
the convexity of the inversion (Section \ref{subsec:Convexity-of-Inversion})
and reasoning about potential biases in the DLS retrievals (Section \ref{subsec:Physical-Consistency-of}). Section \ref{sec:3} proposes the new multi-frequency inversion formalism. Through Monte Carlo simulations, we demonstrate in Section \ref{subsec:A-Monte-Carlo}
that this new approach leads to almost unbiased retrievals with reduced
uncertainty compared to classic unconstrained DLS retrievals. In Section
\ref{subsec:Windowed-Retrievals}, we extend the inversion formalism
for retrievals over a window of time. Section \ref{subsec:Implementation-for-SMAP} discusses implementation of the algorithm for a super-resolved soil moisture retrieval using SMAP data and some initial validation results are presented in Section \ref{sec:validation}. Section \ref{sec:Discussion-and-Concluding} provides discussions, concludes and points out to future directions and the need for a thorough ground validation of the algorithm. 

\section{A review of the $\tau$-$\omega$ Model and its Least-squares Inversion\label{sec:2}}

The $\tau$-$\omega$ model has been explained in numerous seminal
publications \citep[e.g.,][]{Mo1982,Kerr_1990,Njoku1996a}. Here, we briefly discuss
the model to set the stage for investigating its convexity and potential
biases that may arise in its LS inversion. 

The $\tau$-$\omega$ model treats the vegetation layer as a weakly
scattering medium with a low single scattering albedo, typically less
than 0.2 in frequencies from 1 to 5 GHz. The model has three components: (1) emission by the soil surface $(1-r_{p})\gamma_{p}T_{s}$,
(2) upward emission by the slanted column of vegetation with a finite
thickness $(1-\omega_{p})(1-\gamma_{p})T_{c}$, and (3) canopy downward
emission followed by soil coherent reflection $r_{p}(1-\omega_{p})(1-\gamma_{p})\gamma_{p}T_{c}$.
Therefore, the observed brightness $Tb_{p}$ can be expressed as follows:
\begin{equation}
Tb_{p}=(1-r_{pr})\gamma_{p}T_{s}+(1-\omega_{p})(1-\gamma_{p})T_{c}+r_{pr}(1-\omega_{p})(1-\gamma_{p})\gamma_{p}T_{c},\label{eq:1}
\end{equation}
where $T_{s}$ and $T_{c}$ are the effective soil surface and canopy
thermodynamic temperatures, $r_{pr}$ denotes the rough surface soil 
coherent reflectivity, $\gamma_{p}$ is the canopy one-way transmissivity, and
$\omega_{p}$ refers to the vegetation single scattering albedo. Here,
the subscript $p\in\left\{ H,\,V\right\} $ denotes that the quantity
can be horizontally (\emph{H}) or vertically (\emph{V}) polarized.

To make this inversion well-posed, the family of single channel algorithms (SCA) assumes that the $\omega_{p}$ is a known constant for different land surface types and
the vegetation transmissivity is approximately unpolarized and can
be estimated from the NDVI climatology \citep{Jackson1999}. This approach assumes that
the vegetation optical depth $\tau$ is linearly related to the vegetation water content (VWC)  as $\tau=b\,.\text{VWC}$, where $b$ typically varies between 0.05 to 6.0 m\mbox{\textsuperscript{2}~kg
\textsuperscript{-1}}, depending on the vegetation type \citep{Grined_2004}.
The algorithm estimates the VWC from 10-day climatology of NDVI through
some regression equations \citep{Jackson1999,Crow2005,Jackson_2010}
and a-prior knowledge of the plant's stem structure \citep{Hunt1996,Clavo_2008}.
Having the optical depth, the vegetation transmissivity  can be calculated as $\gamma=\text{exp}\left(-\tau\,\sec\phi\right)$,
where  $\phi$  denotes the radiometer incident angle in radians. Then, the smooth
surface soil reflectivity value $r_{p}$ is derived from their rough
counterpart $r_{pr}$ using $r_{pr}=r_{p}\exp\left(-h\,\cos\phi\right)$,
where $h$ is linearly related to the root-mean-squared variability
of the surface height \citep{Choudhury_79}. Finally, the soil moisture
is estimated using the Fresnel equations and a soil dielectric model \citep[e.g.,][]{Dobson1985,Mironov_2009,Mironov2013}.
As is evident, the SCA does not retrieve the vegetation parameters
and can not capture short-term changes in VWC, which could be an important
source of error over grass and croplands. For example, the field campaign Soil
Moisture Experiment 2002 (SMEX02) shows that in a growing season,
the VWC can increase from $\sim0.75$ to 2 km~m\textsuperscript{-2}
every 10 days in a corn field \citep{Jackson2004}. 

The class of dual channel algorithms (DCA) \citep{Njoku1999} often relies on nonlinear LS inversion of the $\tau$-$\omega$ model using
the DLS optimization algorithm. To be more specific, let us assume
that in equation (\ref{eq:1}) the microwave surface emissivity at
polarization $p$ is $e_{p}=Tb_{p}/T_{s}=f(\theta_{p})+\epsilon$,
where the canopy and soil temperature are assumed to be at equilibrium
$T_{s}=T_{c}$, $\theta_{p}=(r_{pr},\,\gamma_{p},\,\omega_{p})^{T}$, and $\epsilon$ denotes an error term that can be approximated well
by the Gaussian distribution. Then, the LS retrieval can be defined
as follows:
\begin{align}
\theta_{p}^{*} & =\underset{\theta_{p}}{\text{argmin}}\,\mathcal{J}(\theta_{p})\nonumber \\
 & =\underset{\theta_{p}}{\text{argmin}}\frac{1}{2}\left(e_{p}-f(\theta_{p})\right)^{2}.\label{eq:2}
\end{align}
Hereafter, we drop the polarization subscript for notational convenience. 

The DLS optimization is a Newton-type algorithm that attempts to construct a sequence $\theta_{k+1}=\theta_{k}+\delta_{k}$ from an initial guess $\theta_{0}$ that converges to $\theta^{*}$, where
$\delta_{k}$  denotes the search direction. In classic Newton's method, the search direction is obtained by minimizing the following quadratic
approximation of the cost function at each iteration:
\begin{equation}
\mathcal{Q}(\delta_{k})=\mathcal{J}(\theta_{k})+\delta_{k}^{T}\,g_{k}+\frac{1}{2}\delta_{k}^{T}H_{k}\delta_{k},\label{eq:3}
\end{equation}
where $g_{k}\stackrel{\text{def}}{=}\nabla\mathcal{J}\left(\theta_{k}\right)$
and $H_{k}$ is a symmetric approximation of the Hessian matrix $\nabla^{2}\mathcal{J}\left(\theta_{k}\right)$
at $k^{\text{th}}$ step. Thus, the search direction can be obtained
by solving $H_{k}\delta_{k}=-g_{k}$ for $\delta_{k}$. This linear
system of equations is underdetermined for inversion of the $\tau$-$\omega$
model for the SMAP single band radiometer.

The classic DLS algorithm approximates the Hessian as $H_{k}\simeq g_{k}\,g_{k}^{T}$
and solves a damped version of the linear system such that $\left(H_{k}+\lambda_{k}I\right)\delta_{k}=-g_{k}$,
where the damping parameter $\lambda_{k}$ is selected adaptively.
Thus, the DLS algorithm uses a Tikhonov regularization
to solve for the search direction, which makes the problem well-posed,
even if the system is underdetermined. Spectral decomposition of $H_{k}=Q_{k}\Lambda_{k}Q_{k}^{T}$
to its eigenvectors in column space of $Q_{k}$ and eigenvalues as
diagonals of $\Lambda_{k}$ leads to $Q_{k}\left(\Lambda_{k}+\lambda_{k}I\right)Q_{k}^{T}\,\delta_{k}=-g_{k}$.
We can see that the search direction depends not only on the gradient
of the cost function but also on the direction and magnitude of the eigenvectors
of the Hessian. This observation prompts us to pose the following
key questions. Is the LS inversion of the $\tau$-$\omega$ model
a convex problem? Are the solutions of the
DLS algorithm physically consistent? 
\begin{figure}[H]
\begin{centering}
\includegraphics[height=6.2cm]{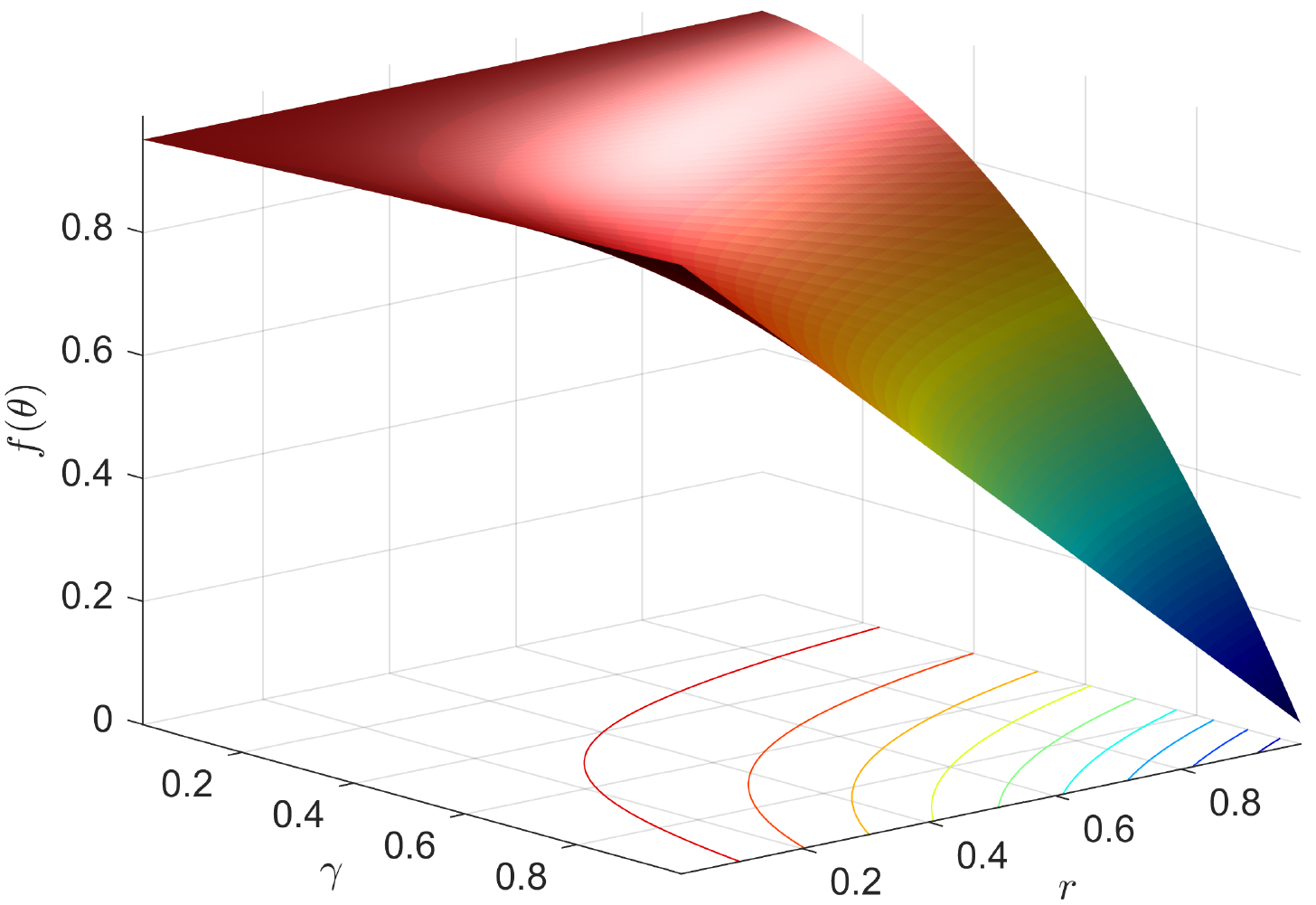}\quad\includegraphics[height=6.2cm]{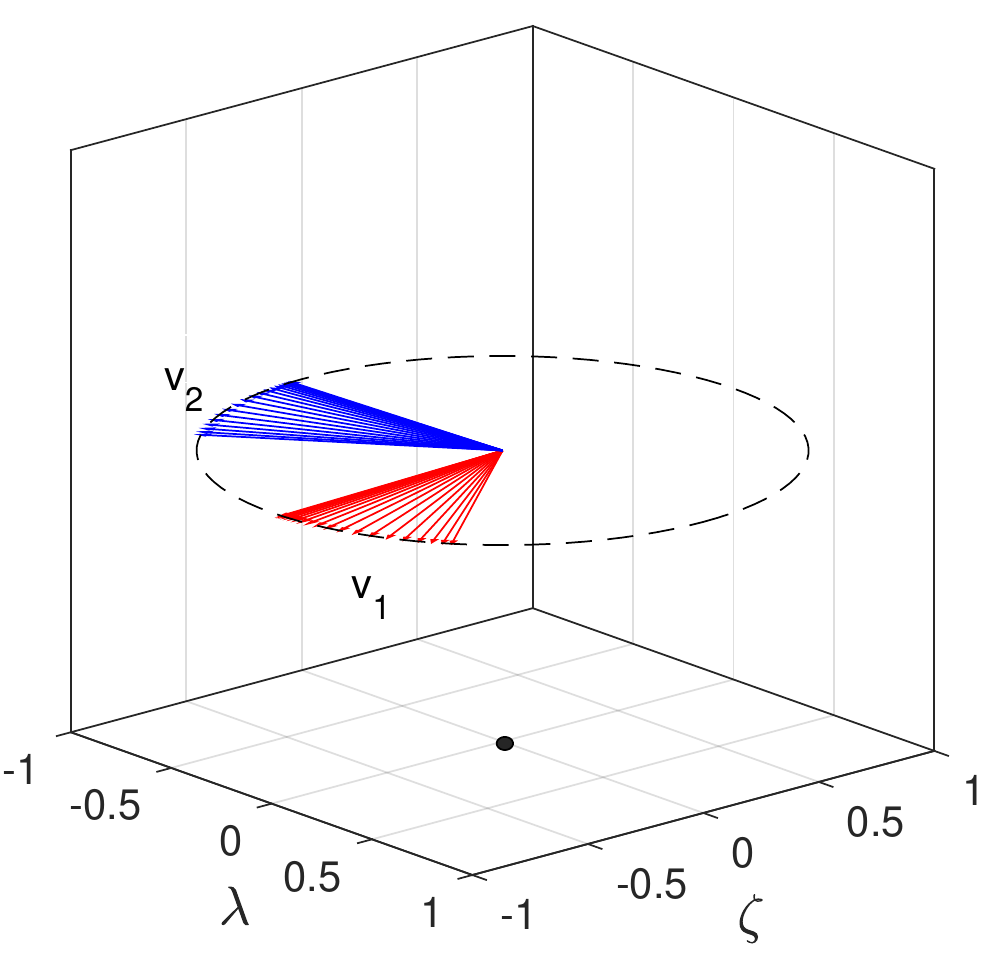}
\par\end{centering}
\caption{The $\tau$-$\omega$ model shows a saddle point
behavior (left) for which the eigenvectors with positive
($\mathbf{v}_{1}$) and negative ($\mathbf{v}_{2}$) eigenvalues (right)
characterize the directions of convexity and concavity of the model.
The eigenvalues are calculated for rough soil surface reflectivity ranging
from 0.4 to 1 and a vegetation optical depth that varies from 0.01
to 3\,m. \label{fig:1} }
\end{figure}
\subsection{Convexity of Inversion\label{subsec:Convexity-of-Inversion}}

The representation of the $\tau$-$\omega$ model through $f(\theta)$ is shown in Figure \ref{fig:1}
for $\omega=0.05$. By visual inspection, we can see that the model
does not produce a strictly convex or concave surface. In effect,
assuming that $\omega$ is a constant, the Hessian is 
\begin{equation}
H^{f}=\left[\begin{array}{cc}
f_{\gamma\gamma} & f_{\gamma r}\\
f_{r\gamma} & f_{rr}
\end{array}\right]=\left[\begin{array}{cc}
2r\left(\omega-1\right) & \zeta\\
\zeta & 0
\end{array}\right],
\end{equation}
where $\zeta=\gamma\left(\omega-1\right)+(\gamma-1)(\omega-1)-1$.
Thus, the Hessian is a saddle point indefinite matrix with
one positive ($\lambda_{1}$) and one negative ($\lambda_{2}$) eigenvalue such that 
$\lambda_{1,\,2}=r(\omega-1)\pm\sqrt{r^{2}\left(\omega-1\right)^{2}+\zeta^{2}}$
 with the corresponding eigenvectors  $\mathbf{v}_{1,\,2}=\left(\lambda_{1,\,2},\,\zeta\right)^{T}$.
These eigenvectors determine two orthogonal directions along which
the model shows convexity and concavity (Figure \ref{fig:1}). Since
$r(\omega-1)<0$, we have $\left|\lambda_{2}\right|\geq\left|\lambda_{1}\right|$ and thus the model is more concave than convex. Assuming that $\omega$ is constant, without loss of generality, we have 

\begin{equation}
\mathcal{J}_{rr}=\frac{\partial^{2}\mathcal{J}}{\partial r^{2}}=\left[\gamma+\gamma(\gamma-1)\left(\omega-1\right)\right]^{2}>0.
\end{equation}

\begin{figure}[H]
\begin{centering}
\includegraphics[height=6.4cm]{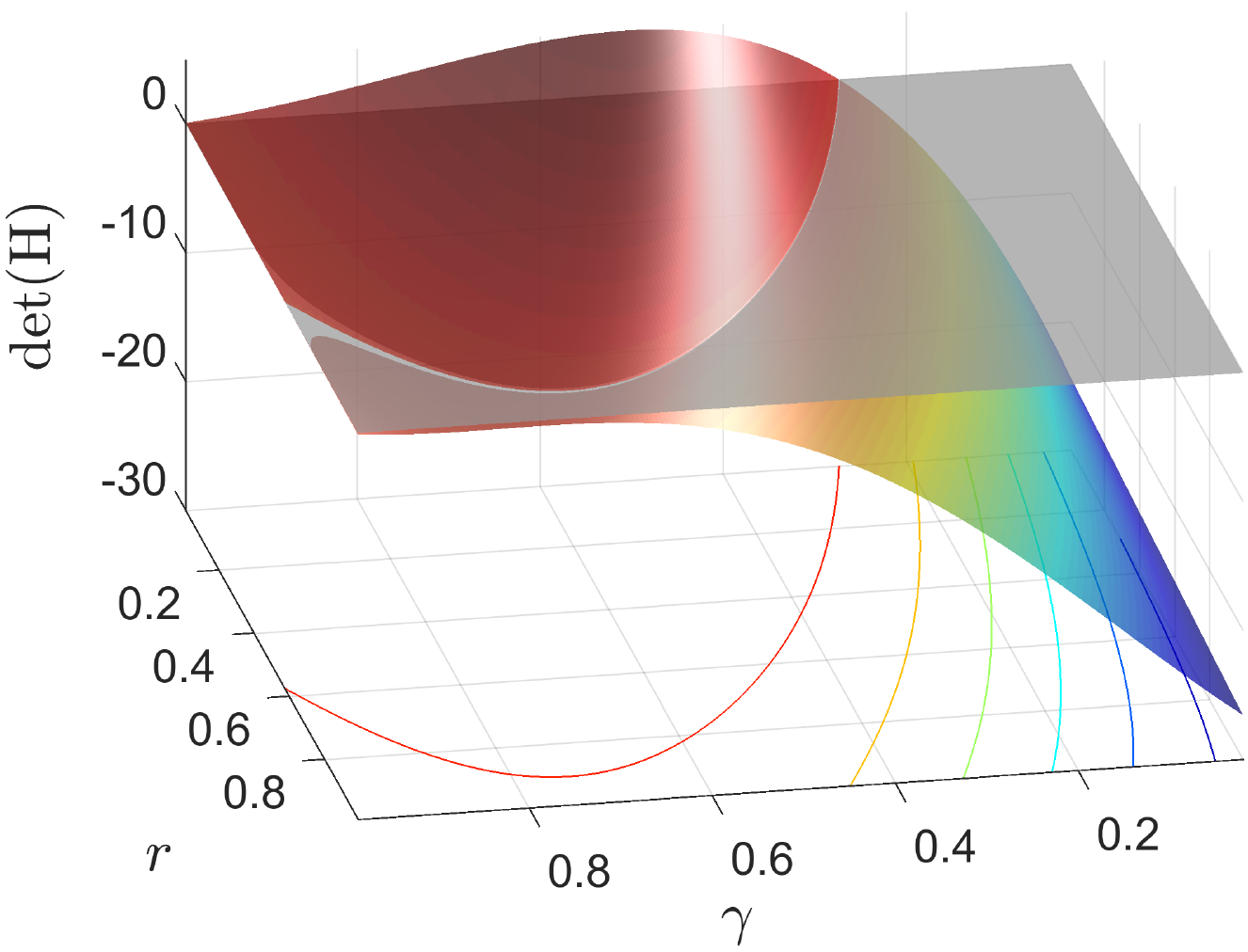}$\quad$\includegraphics[height=6.4cm]{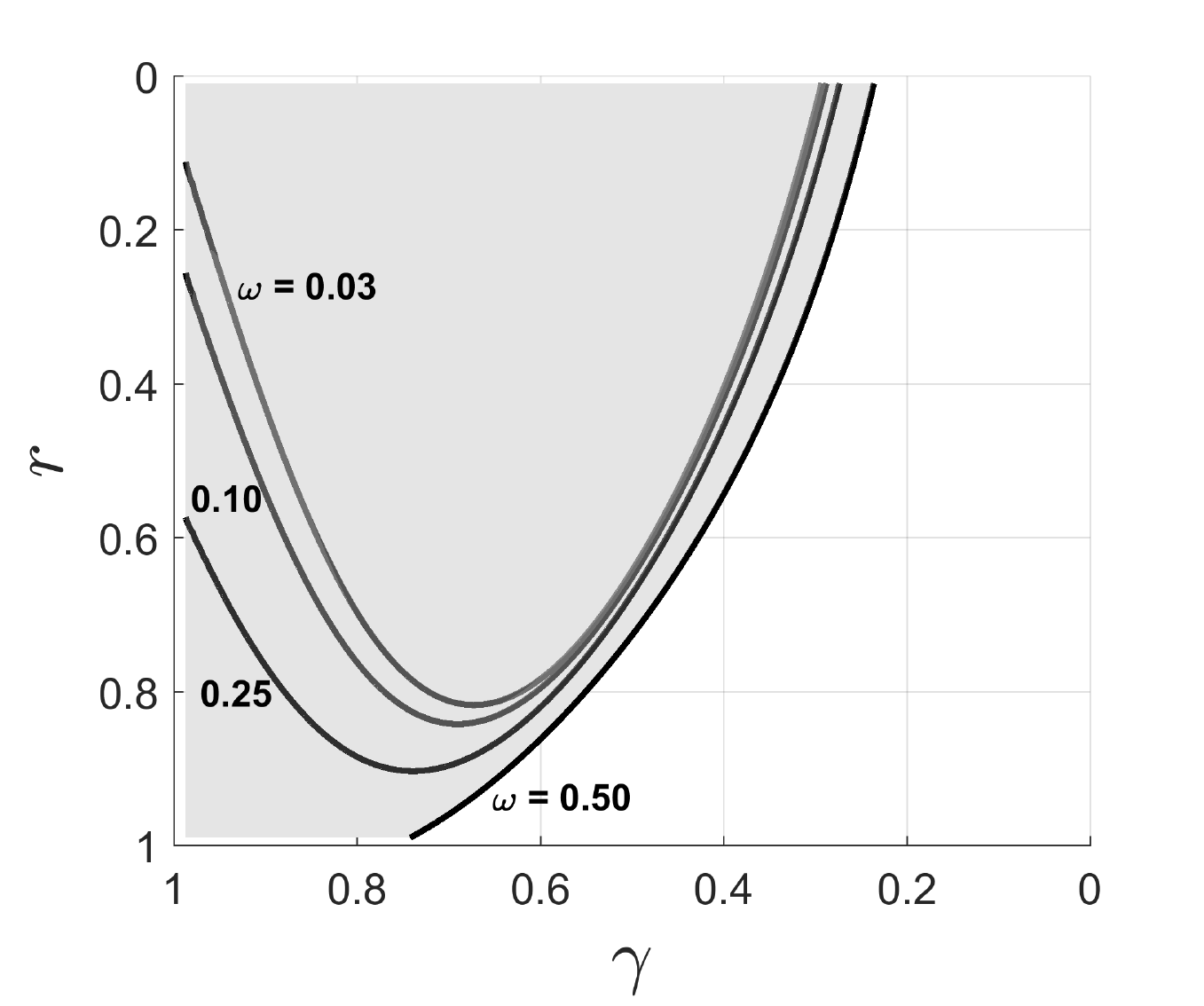}
\par\end{centering}
\caption{Delineating the convex sub-domain of the LS inversion of the $\tau$-$\omega$
model. The shown are the determinant of the Hessian of the LS cost function in problem \ref{eq:2} for $\omega=0.05$ (left) and the shaded
regions over which the Hessian is positive semidefinite (right). \label{fig:2}}
\end{figure}

Therefore,  for characterizing the sub-region over which the problem is convex, it suffices to check where $\det\left(H^{\mathcal J}\right)=\mathcal{J}_{rr}\,\mathcal{J}_{\gamma\gamma}-\left(\mathcal{J}_{r\gamma}\right)^{2}$
is non-negative (Figure \ref{fig:2}). Thus, underdetermined inversion of the $\tau$-$\omega$ model
is not strictly convex all over the feasible domain of the problem.
However, due to monotonic behavior of the model, the LS cost function
exhibits quasi-convexity \citep[p. 98]{Boyd2004}. Due to this non-convexity, the gradient-based approaches such as the DLS algorithm may not converge to the solution curve for those initial points that are outside of the convex domain of the problem. 

\subsection{Physical Consistency of the Inversion\label{subsec:Physical-Consistency-of}}



\begin{figure}[h]
\begin{centering}
\includegraphics[height=7.5cm]{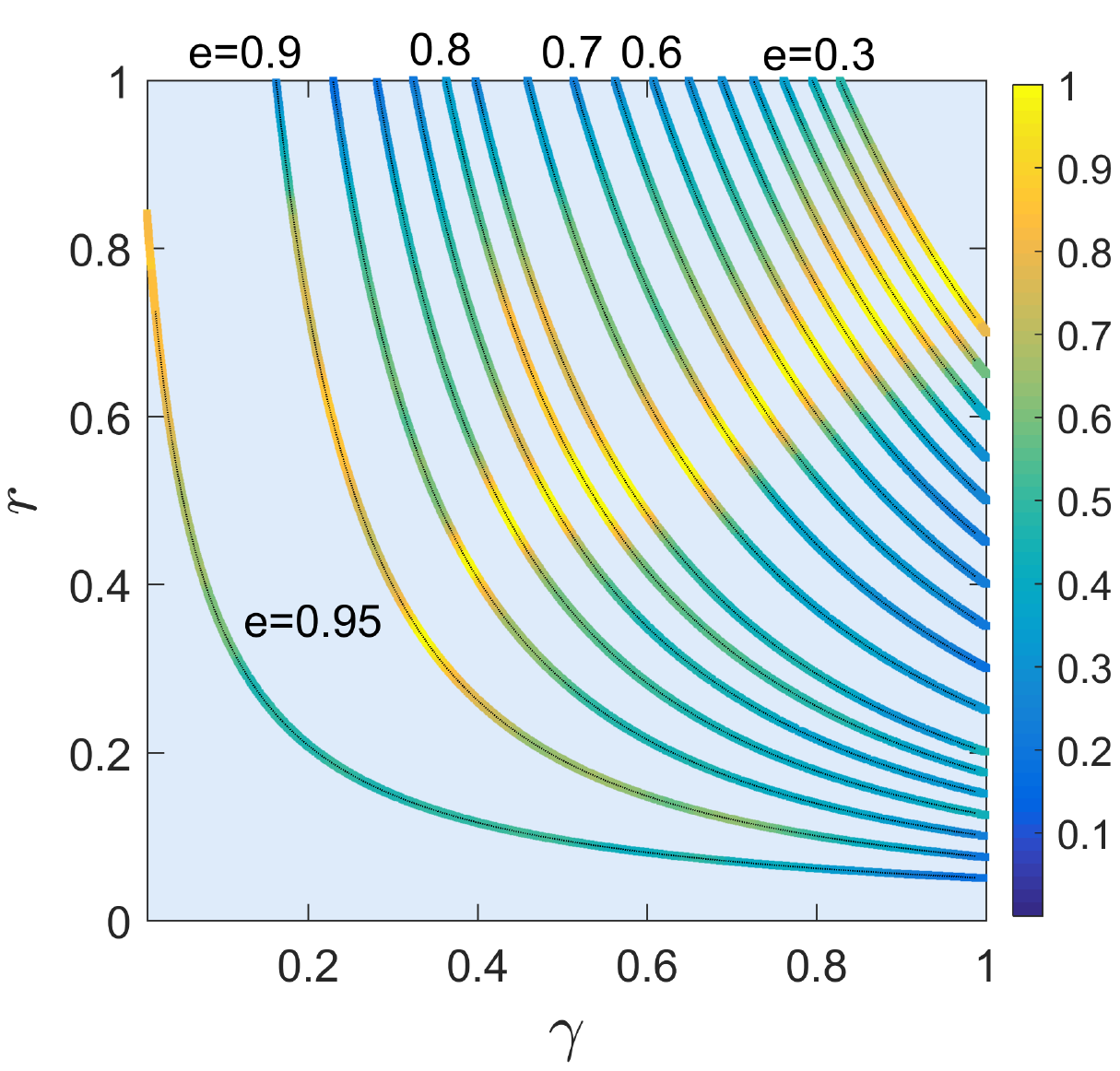}$\qquad$\includegraphics[height=7.5cm]{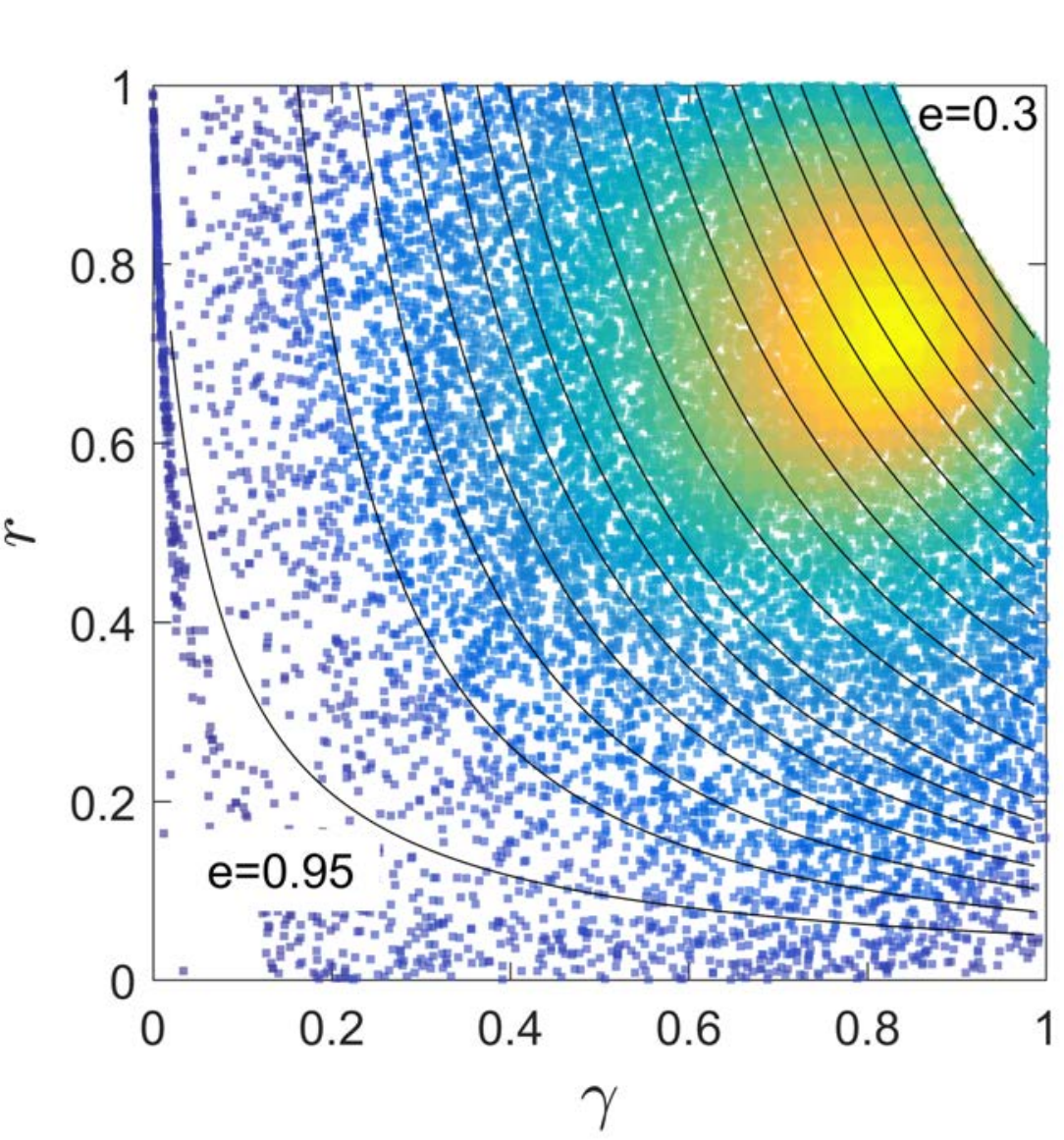}
\par\end{centering}
\caption{The density of retrievals showing preferential convergence pathways of the DLS solutions. The density of the retrievals is obtained through $2.5$e+4 random simulations, uniformly distributed over a few level sets (left) and the entire feasible space of the soil surface reflectivity and vegetation transmissivity values (right). The non-uniform distribution of solutions demonstrates that the simultaneous retrievals could systematically overestimate (underestimate) the soil moisture and VWC when the soil is dry (wet) under a sparse (dense) vegetation. \label{fig:3}}
\end{figure}

As discussed, the search direction of the DLS retrievals could follow preferential paths depending on the direction of the gradient and eigenvalues of the Hessian. The question is \textendash how do these preferential search paths affect the solution space and physical consistency of the soil moisture retrievals? To answer this question, we conducted Monte Carlo retrieval experiments for two different scenarios. First, we conduced the retrievals for a few values of microwave surface emissivity ranging from 0.3 to 0.95, while the initial soil reflectivity and vegetation transmissivity are randomly drawn between 0 and 1 from a uniform distribution. Second, the input microwave emissivity values are also randomized by drawing samples from a uniform distribution between 0 and 1. To run the DLS in both cases, we consider $\lambda_{0}=0.01$ and $\lambda_{k+1}=0.1\,\lambda_{k}$, when the search direction reduces the cost function, otherwise $\lambda_{k+1}=10\,\lambda_{k}$. The density scatter plot of both experiments are shown in Figure \ref{fig:3}.

The results reveal a non-uniform solution space for the DLS algorithm.
The retrievals are mostly concentrated on the upper right corner of the
feasible solution space and mostly around the mid-point of the level
sets. Concentration of the retrievals near the upper corner of the
plot is not surprising because the feasible solution curves are shorter
for higher values of $r$ and $\gamma$ and thus a discrete approximation of the solution space should be denser over that region. However, concentration of the solutions around the mid-point of the level sets shows the effects of preferential descending paths, which can lead to biased and physically inconsistent retrievals.

Overall, it seems that for sparse (dense) vegetation with a
shallow (deep) optical depth, the DLS tends to underestimate (overestimate) the transmissivity and overestimate (underestimate) the surface reflectivity, which results in overestimation (underestimation) of both soil moisture and VWC. This pattern of biased retrievals is more pronounced when the microwave emissivity is below 0.9. For dry soil with higher microwave emissivity, it appears that the retrievals tend to systematically overestimate both of the variables.

\section{Constrained Inversion of the $\tau$-$\omega$ Model\label{sec:3}}

The systematic biases could be reduced by constraining
the solution space using an a priori knowledge about the feasible
range of the soil moisture and VWC \citep{Wigneron2000}. In reality, the soil moisture and VWC are not unbounded physical parameters. In any retrieval scene, the soil moisture is bounded by soil porosity and often varies between the the permanent wilting (PWP) point and natural saturation. The VWC and its temporal dynamics also largely depend on the plants physiology and phenology. To
avoid retrieval biases, we suggest the following constrained retrieval
approach:
\begin{equation}
\theta^{*}=\underset{\theta}{\text{argmin}}\frac{1}{2}\left(e-f(\theta)\right)^{2}\,\,\,\,\text{subject to}\,\,\,\,\theta_{l}\preceq\theta\preceq\theta_{u},\label{eq:6}
\end{equation}
where $\theta_{l}=(r_{l},\,\gamma_{l})^{T}$, $\preceq$ is an element-wise inequality and $\theta_{u}=(r_{u},\,\gamma_{u})^{T}$
denotes the lower and upper bounds of the input parameters. This problem is not tractable unless we provide additional a priori knowledge that turns it into an (over)determined problem. 

The class of constrained nonlinear LS
problems is often solved by a family of optimization techniques called
the Trust Region (TR) algorithms \citep{Goldfeld1966,Sorensen1982}.
In this method
the search direction at each iterate is obtained by minimizing a quadratic
approximation of the objective function over a restricted ellipsoidal
region centered on the current iterate. Therefore, at each step, the algorithm solves a constrained quadratic sub-problem, which makes it computationally more burdensome than the DLS approach. However, this construction prevents overshooting the local minima and thus could improve the convergence rate. The constrained sub-problem can be further equipped with computationally cheap projection operators to map the solution, at each iterate $\theta_{k+1}=\theta_{k}+\delta_{k}$, onto the convex set of $\theta_{l}\preceq\theta_{k}\preceq\theta_{u}$
\citep{Coleman1999}. 

To  recast the inversion to an overdetermined problem, we suggest to add a Tikhonov regularization  as follows:
\begin{equation}
\theta^{*}=\underset{\theta}{\text{argmin}\,}\left(e-f(\theta)\right)^{2}+\lambda\left\Vert \theta\right\Vert _{2}^{2}\,\,\,\,\text{subject to}\,\,\,\,\theta_{l}\preceq\theta\preceq\theta_{u}.\label{eq:7}
\end{equation}
where $\lambda>0$ is a non-negative parameter and $\left\Vert \theta\right\Vert _{2}^{2}$ is the 2-norm or sum of the squares of the unknownparameters. The above problem can be recast to a standard form as follows:
\begin{equation}
\theta^{*}=\underset{\theta}{\text{argmin}}\,\left\Vert \left(\begin{array}{c}
e-f(\theta)\\
\lambda^{1/2}\theta
\end{array}\right)\right\Vert _{2}^{2}\,\,\,\,\text{subject to}\,\,\,\,\theta_{l}\preceq\theta\preceq\theta_{u}.\label{eq:8}
\end{equation}

\begin{figure}[t]
\begin{centering}
\includegraphics[width=0.5\textwidth]{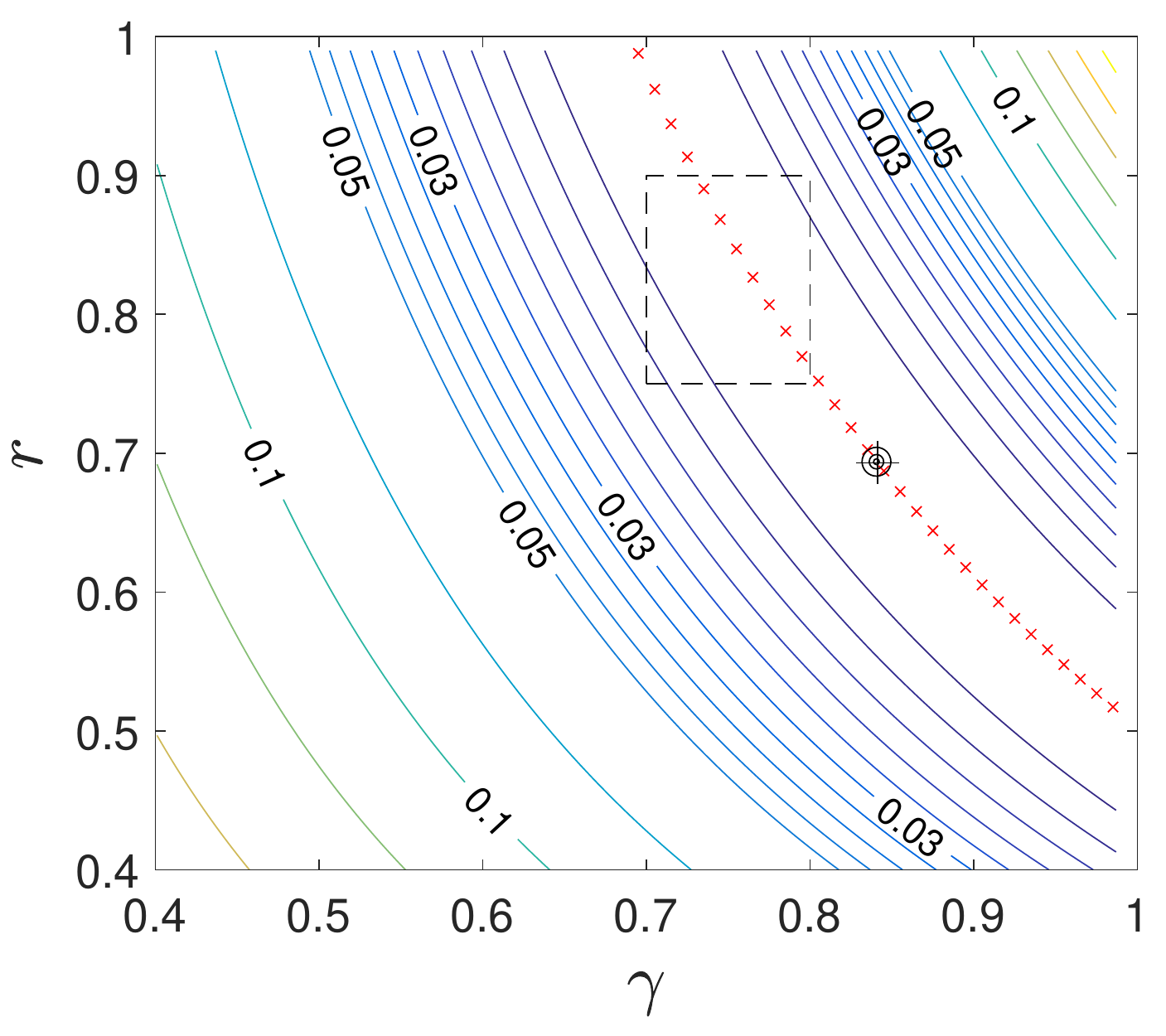}\includegraphics[width=0.5\textwidth]{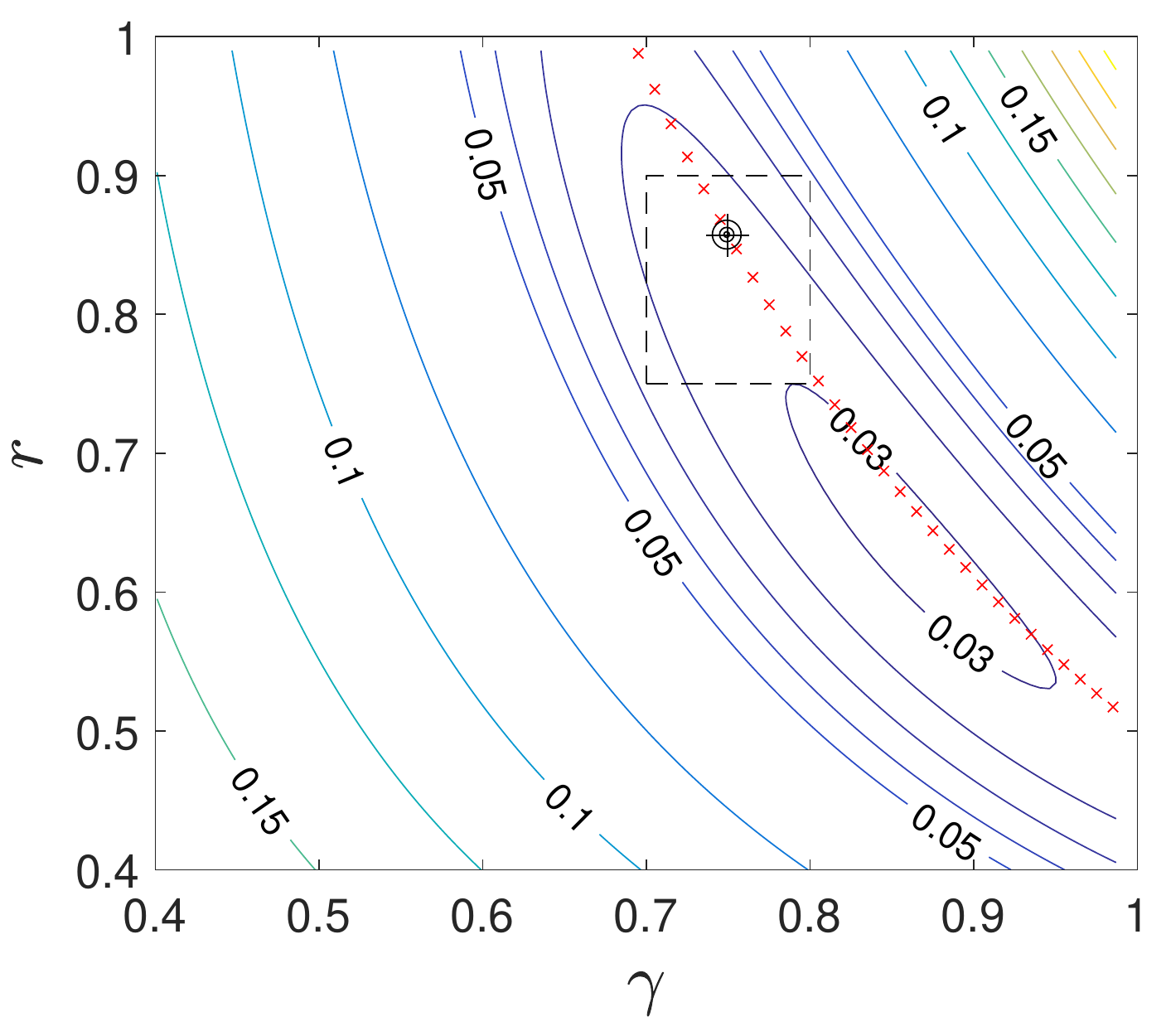}
\par\end{centering}

\caption{The level sets of the cost function in problem \ref{eq:8}
for inversion of the $\tau$-$\omega$ model, with $\lambda=0.0$ (left) and $\lambda=0.015$ (right), in which $T_{s}=298$ K, $Tb=148$ K, and $\omega=0.05$. The solution curve
for $\lambda=0$ is shown with red crosses. The black concentric circles with the plus sign
at the center show the solution of unconstrained (left, equation \ref{eq:2})
and constrained retrievals (right, equation \ref{eq:8}). The constrained
retrieval is obtained assuming that $\gamma\in\left[0.70\,0.80\right]$
and $r\in\left[0.75\,0.90\right]$\textemdash shown by a rectangle
with dashed sides.
\label{fig:4}}
\end{figure}

For ill-posed linear inverse problems, it is well documented that
the Tikhonov regularization leads to a unique solution with reduced uncertainty. This reduced uncertainty often comes at the expense of a bias in the solution \citep{Hansen2010} that can be
negligible for sufficiently small values of $\lambda$. Figure \ref{fig:4}
shows the level sets of the cost function in problem \ref{eq:8} for
$\lambda=0$ and 0.015, when $\omega=0.05$. When $\lambda\rightarrow0$
, the problem approaches to the classic LS inversion with infinite
number of solutions. For positive values of $\lambda$, a convex region
is formed, which narrows down the solution space and reduces the retrieval uncertainties.

To expand the above inversion to a multi-channel algorithm, let us assume that $\mathbf{e}=\left(e_{1},\ldots,e_{q}\right)^{T}\in\mathbb{R}^{q}$
and $\boldsymbol{\theta}=\left(\theta_{1},\ldots,\theta_{q}\right)^{T}\in\mathbb{R}^{nq}$
denote the microwave emissivity and input parameters at $q$ frequencies,
where $\theta_{q}\in\mathbb{R}^{1\times n}$ denotes a column vector with $n$ unknown
free parameters. In addition, one may want to consider a $q$-by-$q$
error covariance $\mathbf{E}\in\mathbb{R}^{q\times q}$ to
account for the precision of each channel. In this setting a multi-channel
formulation of the above inversion scheme can be represented in the
following standard form:

\begin{equation}
\boldsymbol{\theta}^{*}=\underset{\boldsymbol{\theta}}{\text{argmin}}\,\left\Vert \left(\begin{array}{c}
\mathbf{E}^{-1/2}\left(\mathbf{e}-f(\boldsymbol{\theta})\right)\\
\lambda^{1/2}\boldsymbol{\theta}
\end{array}\right)\right\Vert _{2}^{2}\,\,\,\,\text{subject to}\,\,\,\,\boldsymbol{\theta}_{l}\preceq\boldsymbol{\theta}\preceq\boldsymbol{\theta}_{u},
\end{equation}

\noindent where $f(\boldsymbol{\theta}):\,\mathbb{R}^{nq}\rightarrow\mathbb{R}^{q}$
and there are $(n+1)q$ equations and $nq$ unknowns. 

Additionally, extension to the retrievals over a window of time can be recast as
follows:

\begin{equation}
\boldsymbol{\theta}_{\tau}^{*}=\underset{\boldsymbol{\theta}_{\tau}}{\text{argmin}}\left\Vert \left(\begin{array}{c}
\mathbf{\mathbf{E}}_{\tau}^{-1/2}\left(\mathbf{e}_{\tau}-f(\boldsymbol{\theta}_{\tau})\right)\\
\Lambda^{1/2}\mathbf{D}\boldsymbol{\theta}_{\tau}
\end{array}\right)\right\Vert _{2}^{2}\,\,\,\,\text{subject to}\,\,\,\,\boldsymbol{\theta}_{l\tau}\preceq\boldsymbol{\theta}_{\tau}\preceq\boldsymbol{\theta}_{u\tau},\label{eq:12}
\end{equation}
where $f(\boldsymbol{\theta}_{\tau}):\,\mathbb{R}^{nqt}\rightarrow\mathbb{R}^{qt}$
, $\mathbf{e}_{\tau}=\left(\mathbf{e}_{1}^{T},\ldots,\mathbf{e}_{t}^{T}\right)^{T}\in\mathbb{R}^{qt}$
and $\boldsymbol{\theta}_{\tau}=\left(\mathbf{\boldsymbol{\theta}}_{1}^{T},\ldots,\mathbf{\boldsymbol{\theta}}_{t}^{T}\right)^{T}\in\mathbb{R}^{nqt}$
stack all the observed microwave emissivity values and free parameters
in a vector form over a window of time with size $\tau=1,\ldots,t$. Here, $\mathfrak{\mathbf{E}}_{\tau}\in\mathbb{R}^{qt\times qt}$
is a block diagonal matrix in which, each block contains the channel
error covariance matrix $\mathbf{E}\in\mathbb{R}^{q\times q}$ and
$\mathbf{D}\in\mathbb{R}^{m\times nqt}$ ($m\leq nqt$) is a linear
transformation that can impose an a priori information about the temporal
variability of the free parameters. For example, instead of assuming
that the vegetation water content remains constant over a window of
time, one may assume that it is a slow varying process. One way to
formalize this assumption is to impose a certain degree of smoothness
by minimizing the variance of a temporal derivative of the vegetation
transmissivity. As a result, to properly scale the problem, we might need different regularization
parameters for the soil reflectivity and vegetation transmissivity, which are encoded by the diagonal matrix $\Lambda=\textrm{diag}\left(\lambda_1,\ldots,\lambda_m\right)\in\mathbb{R}^{m\times m}$. Hereafter, we refer to the presented approach as the Constrained Multi-Channel
Algorithm (CMCA).

\section{Implementation, Results and Validation}

In this section, we implement and validate the performance of the
CMCA algorithm in three steps. In the first step, we examine the
results using a Monte Carlo approach to compare the CMCA performance
with unconstrained DLS inversion and evaluate its results under fully
controlled boundary conditions. In the second step, we use soil moisture
gauge data to evaluate the results of the CMCA algorithm for retrievals
over a window of time. In the third step, we implement the algorithmic
for the SMAP satellite observations and validate it against soil moisture gauge observations.

\subsection{A Monte Carlo Validation \label{subsec:A-Monte-Carlo}}

For conducting a controlled validation, we adopt a Monte
Carlo approach. To that end, we generate a statistically representative
number of random combinations of physically feasible free input parameters and simulate their brightness temperatures using the forward $\tau$-$\omega$
model at 1.4 GHz and incident angle 40$^{\circ}$. These brightness
temperatures are then used for retrievals of the known free parameters.
In the experiments, we assume that the rough surface soil reflectivity
values are polarization dependent while vegetation transmissivity
is not. 


Throughout, to confine the computational domain, we assume that $\omega=0.05$,
the constant that relates VWC to its optical depth  $b=0.10$, and
the soil surface roughness parameter $h=0.12$ \citep[see ][]{Choudhury_79}.
To understand the effects of vegetation density and soil types on
the accuracy of retrievals, we conduct our experiments separately for three ranges
of VWC between 0 and 1.5, 1.5 and 3.0, and 3.0 and 5.0 kg~m\textsuperscript{-2}.
The simulations are also stratified based on the National Resources
Conservation Service (NRCS) soil texture classification. 
In the experiments, we assume that the soil moisture varies between
the permanent wilting point (suction head -1500 kPa) and the field
capacity (suction head -33 kPa). The bounds of the rough soil surface reflectivity values for the NRCS soil types are reported in Table \ref{tab:1}.
\begin{table} \scriptsize
\begin{centering}
\begin{tabular}{|l|c|c|c|c|c|c|c|}
\hline 
{Texture} & {PWP (\%)} & {FC(\%)} & {Clay content} & {$r_{Hr\ell}$} & {$r_{Hru}$} & {$r_{Vr\ell}$} & {$r_{Vru}$}\tabularnewline
\hline 
\hline 
{Clay} & {30.0} & {42.0} & {40.0-100} & {0.27} & {0.50} & {0.11} & {0.30}\tabularnewline
\hline 
{Silty clay} & {27.0} & {41.0} & {40.0-60.0} & {0.32} & {0.48} & {0.15} & {0.30}\tabularnewline
\hline 
{Silty clay loam} & {22.0} & {38.0} & {27.5-40.0} & {0.32} & {0.48} & {0.15} & {0.30}\tabularnewline
\hline 
{Clay loam} & {22.0} & {36.0} & {27.5-40.0} & {0.32} & {0.48} & {0.15} & {0.30}\tabularnewline
\hline 
{Silt} & {6.00} & {30.0} & {0.00-12.5} & {0.16} & {0.45} & {0.05} & {0.27}\tabularnewline
\hline 
{Silt loam} & {11.0} & {31.0} & {0.00-27.5} & {0.20} & {0.46} & {0.07} & {0.28}\tabularnewline
\hline 
{Sandy clay} & {25.0} & {36.0} & {35.00-55.0} & {0.31} & {0.46} & {0.15} & {0.28}\tabularnewline
\hline 
{Loam} & {14.0} & {28.0} & {7.50-27.50} & {0.25} & {0.43} & {0.10} & {0.25}\tabularnewline
\hline 
{Sandy clay loam} & {17.0} & {27.0} & {20.0-35.0} & {0.27} & {0.40} & {0.11} & {0.23}\tabularnewline
\hline 
{Sandy loam} & {8.00} & {18.0} & {0.00-20.0} & {0.18} & {0.35} & {0.06} & {0.18}\tabularnewline
\hline 
{Loamy sand} & {5.00} & {12.0} & {0.00-15.0} & {0.15} & {0.28} & {0.04} & {0.12}\tabularnewline
\hline 
{Sand} & {5.00} & {10.0} & {0.00-10.0} & {0.16} & {0.25} & {0.04} & {0.10}\tabularnewline
\hline 
{\footnotesize{}All soil types} & {-} & {-} & {-} & {0.15} & {0.50} & {0.04} & {0.30}\tabularnewline
\hline 
\end{tabular}
\par\end{centering}{\small \par}
\caption{The lower and upper bounds of the soil moisture and the rough soil
surface reflectivity values at horizontal ($r_{Hr}$) and vertical
polarization ($r_{Vr}$) for different NRCS soil types at incident
angle 40$^{\circ}$ and $f=1.4$ GHz, using the soil dielectric model by \citet{Mironov_2009}. The values of the permanent wilting point
(PWP) and the field capacity (FC) are obtained from \citep{Saxon2006}.
\label{tab:1}}
\end{table}

In Table \ref{tab:1}, silt and sand show the widest and narrowest
range of surface soil surface reflectivity, respectively\textemdash largely because of the range of their water capacity. It is important to note that,
based on the chosen dielectric model, the variability range of $r_{Hr}$
is $\sim50$\% larger than its vertical counterpart\textemdash when
the real part of the soil dielectric constant varies between 0 and
20. Generally speaking, a wider range of surface reflectivity may imply more sensitivity of the channel to the changes of soil moisture content. However, provided that the bounds are physically consistent, a narrower bound is a stronger
prior, which could lead to reduced retrieval uncertainties. Certainly, this argument shall be interpreted independent of the observation noise and accuracy
of the radiometer at different polarization channels. 

To evaluate the result of the CMCA, we first need to determine an
optimal value of the regularization parameter $\lambda$ in equation
\ref{eq:8}, for which there is no closed form expression. 

\begin{figure}
\begin{centering}
\includegraphics[width=1\textwidth]{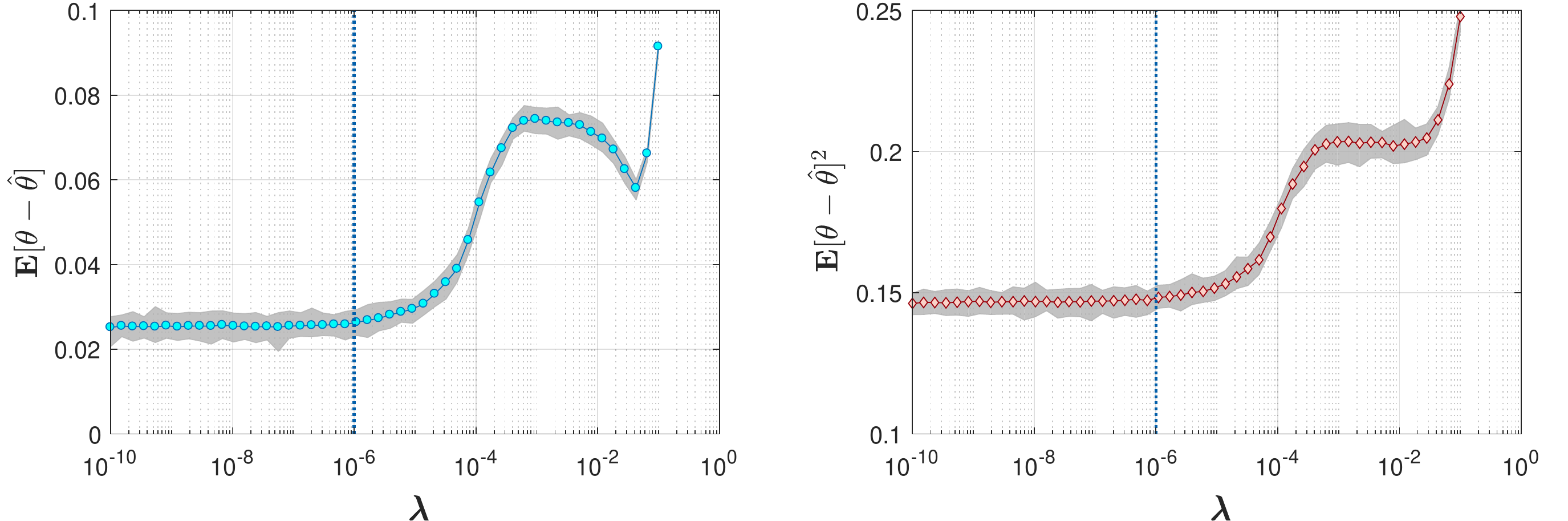}
\par\end{centering}

\caption{The bias (left) and the RMSE (right) of the CMCA retrievals
for 60 retrieval ensembles (shaded region) as a function of the regularization
parameter $\lambda$ ranging from 1e-10 to 1, where $\theta=\left(r_{Hr},\,r_{Vr},\,\gamma\right)^{T}$. \label{fig:5}}
\end{figure}

Figure \ref{fig:5} shows the bias and root mean squared error (RMSE)
of CMCA retrievals as a function of $\lambda$, where $\theta=\left(r_{rH},\,r_{rV},\,\gamma\right)^{T}$.
The shaded region shows the upper and lower bounds of 60 ensemble
simulations. Specifically, we generated 2.5e+3 uniformly
distributed random inputs of surface temperature 0\textendash 40$^{\circ}$,
soil moisture 0.05\textendash 0.42, VWC 0\textendash 5.0 kg~m\textsuperscript{-2},
and soil clay content 0\textendash 99\%. Based on these input parameters,
we ran the $\tau$-$\omega$ model and corrupted the simulated brightness
temperatures with a zero-mean white Gaussian noise with standard deviation
1.3\,K to resemble the observed brightness temperatures by the
SMAP radiometer. Then, the bias and RMSE of the CMCA retrievals are
obtained as a function of $\lambda$. We repeated this process for
60 times to evaluate the sensitivity of the retrievals to the observation noise. As is evident, for $\lambda$ values $\leq$\,1e-6, 
the bias and RMSE are minimum and the solutions are fairly stable.
Above this threshold the the quality metrics begin to grow rapidly.
As the $\lambda $ becomes larger, the shaded areas begin to gradually shrink,
which means the observation noise is further suppressed and the solutions
become more stable. However, we can see that the bias is also growing. 

\begin{figure}
\begin{centering}
\includegraphics[width=1\textwidth]{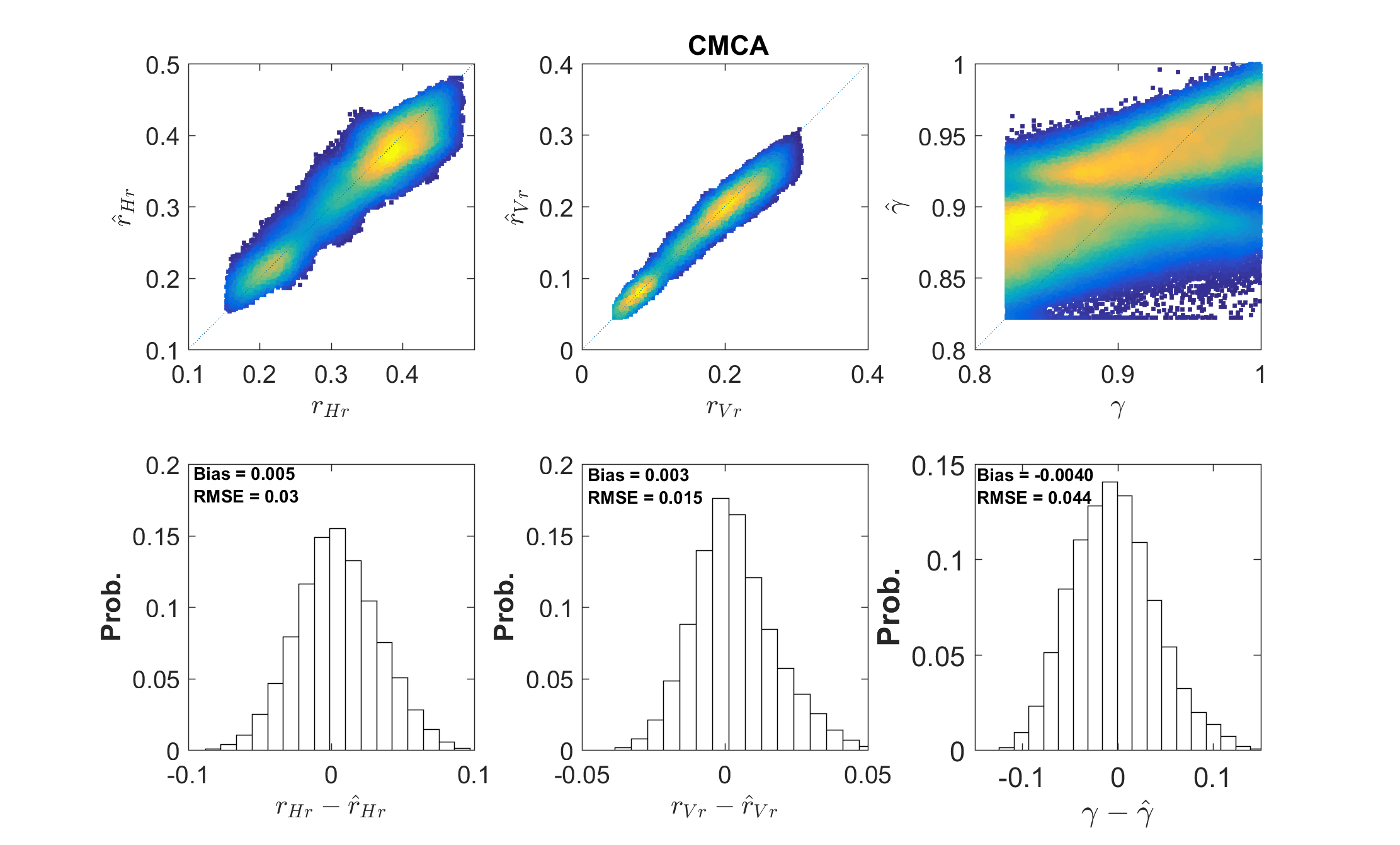}
\includegraphics[width=1\textwidth]{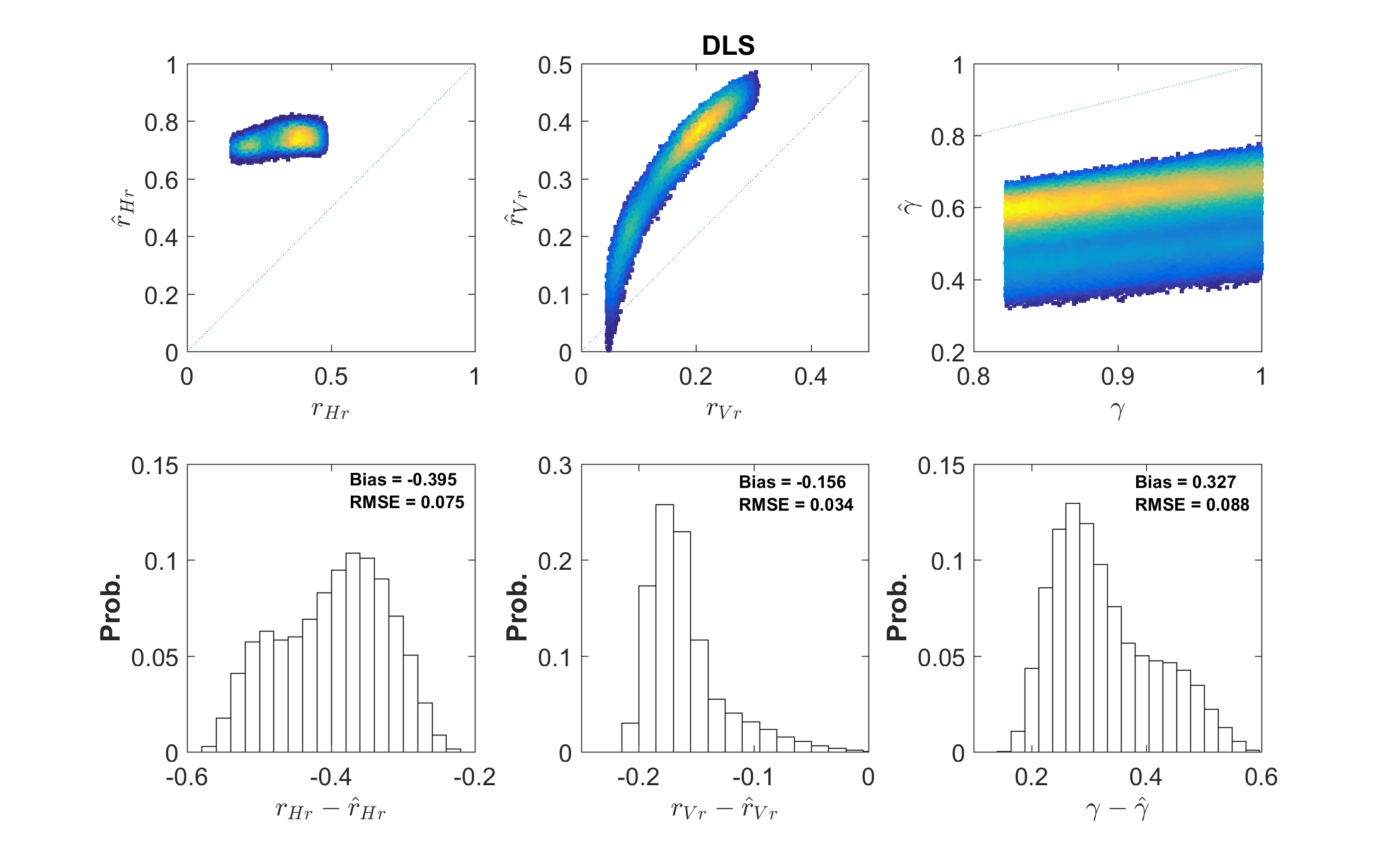}
\par\end{centering}

\caption{Results of CMCA (top two rows) versus unconstrained DLS (bottom two rows) retrievals of the rough soil reflectivity at horizontal ($r_{Hr}$) and vertical ($r_{Vr}$) polarization and the vegetation transmissivity $\gamma$, obtained via a controlled Monte Carlo experiment. \label{fig:6}}
\end{figure}
Figure \ref{fig:6} compares the results of CMCA with DLS solutions for $0<\text{VWC}<1.5$
kg~m\textsuperscript{-2} over all soil types for 5e+5 random inversion
scenarios, where both channels are considered to be equally important.
We first found that for the unconstrained DLS, involving both channels independently,
may not necessarily decrease the retrieval error due to the ill-conditioning of the inversion, which often renders the information content of one channel ineffective.
For example in Figure \ref{fig:6} (third row, first column), we can
see that the retrieved $\hat{r}_{rH}$ through the DLS approach does
not contain meaningful information. However, because of the used Tikhonov
regularization in CMCA, the condition number of the problem is increased and the retrievals of the reflectivity values at both channels are correlated well with the reference.

The results in Figure \ref{fig:6} verify our earlier finding that
the unconstrained underdetermined inversion of the $\tau$-$\omega$
model could lead to biased results. It is shown that for high values
of vegetation transmissivity, the soil moisture and VWC are likely
to be overestimated by DCA. We need to note that with a proper bias correction,
the RMSE values are in an admissible range. For example, we can see
that the bias and RMSE in DCA retrievals of $r_{Vr}$ are  -0.16
and +0.034, respectively, which are around 50 and 11\% of its feasible range of variability. 

\begin{figure}[H]
\begin{centering}
\includegraphics[width=1\textwidth]{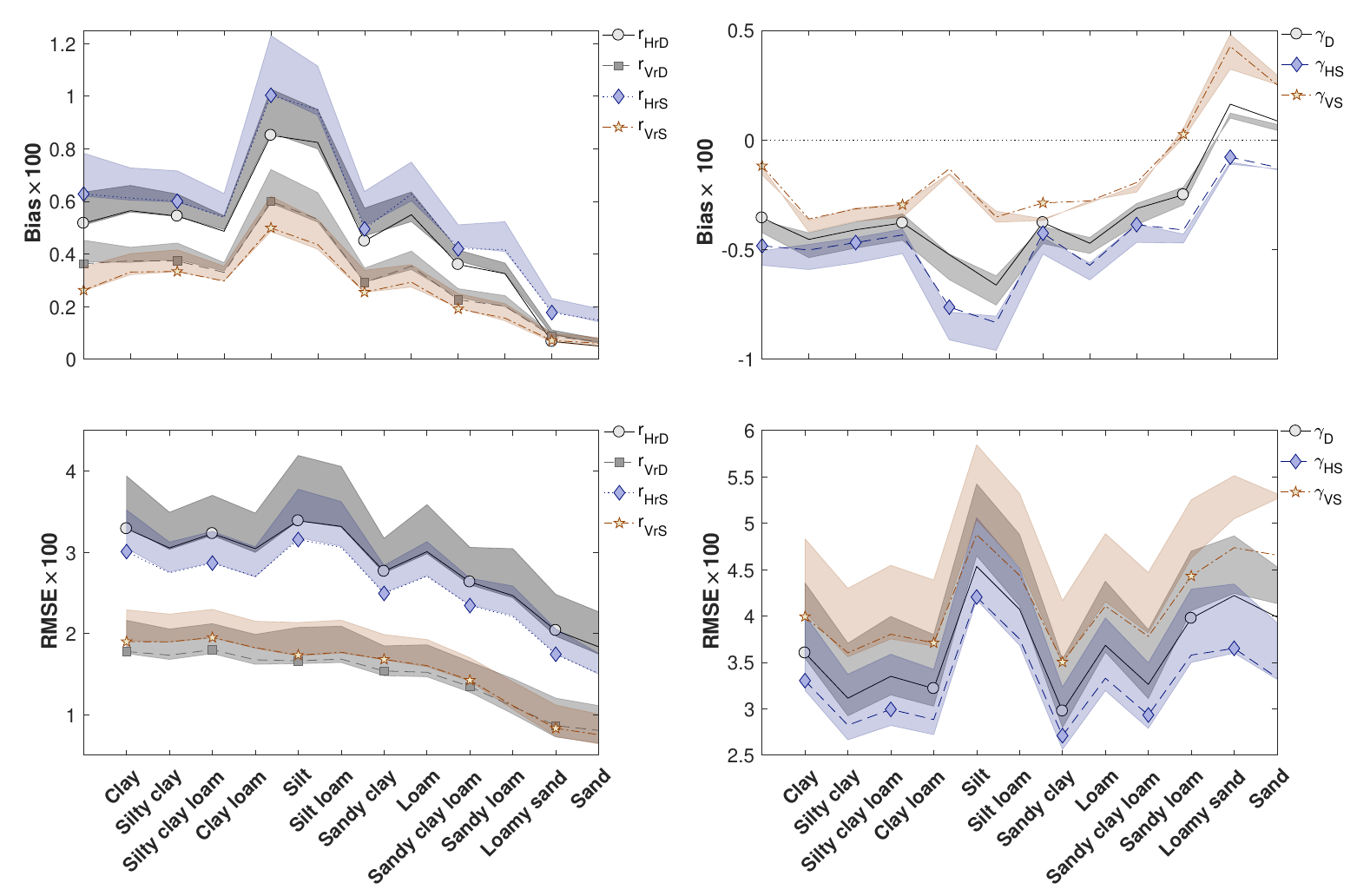}
\par\end{centering}

\caption{The normalized quality metrics by their feasible range for the CMCA retrievals over different soil types and VWC. The results are shown at horizontally ($H$) and vertically ($V)$ polarizations for the dual ($D$) and single ($S$) channel retrievals.
The lines with markers show the results for VWC ranging from 1.5 to 3.0
kg\,m\protect\textsuperscript{-2}, while the upper and lower bounds
of the shaded areas show the results for higher 
(3.0\textendash 5.0 kg\,m\protect\textsuperscript{-2}) and lower
(0\textendash 1.5 kg\,m\protect\textsuperscript{-2}) ranges of VWC. \label{fig:7}}
\end{figure}

We extend the above analysis to different NRCS soil types (Figure \ref{fig:7}). To that end, the biases and RMSE values are obtained for 5e+5 randomized retrievals for each soil type over three ranges of VWC. Overall, the results are almost unbiased and the RMSE is in a reasonable range. For almost all soil types, the  bias (RMSE) remains below 5 and 1\% (25 and 35\%) in retrievals of the soil reflectivity and vegetation transmissivity, respectively. 

In the first column of Figure \ref{fig:7}, we can see that both bias and RMSE of the retrieved reflectivity values are improved as
the soil clay content is decreased. This observation implies that the quality
of the soil moisture retrievals highly depends on proper determination
of the reflectivity bounds, which are tighter for sandy soils and
wider for silty soils. We also observe that the retrievals normally
tend to systematically overestimate the surface soil reflectivity,
even though the biases are sufficiently small. The quality of both
dual and single channel retrievals are also examined and it is found
that the uncertainties are reduced when the vertical channel is involved,
which can be due to its tighter bounds. We also found that the quality
of the retrieved reflectivity values is not excessively sensitive
to the VWC when it remains below 3 kg\,m\textsuperscript{-2}; however,
the bias and RMSE decrease by 20\% and 16\%\textemdash when
VWC increases from 3.0 to 5.0 kg\,m\textsuperscript{-2}, respectively.

In the second column of Figure \ref{fig:7}, the quality
metrics for retrievals of the vegetation transmissivity ($\gamma$)
are shown, which do not exhibit any significant trend as a function of soil type. However, the metrics are slightly increased over silty soils\textendash mainly because retrievals of $\gamma$ and soil reflectivity values are not independent. The retrievals are almost unbiased; however, tend to
systemically underestimate $\gamma$, except for soils with high sand
content. The retrievals in vertical polarization provide minimum biases
but the horizontal channel leads to smaller RMSE values.

\subsection{Windowed Retrievals \label{subsec:Windowed-Retrievals}}

\begin{figure}[t]
\begin{centering}
\includegraphics[width=1.0\textwidth]{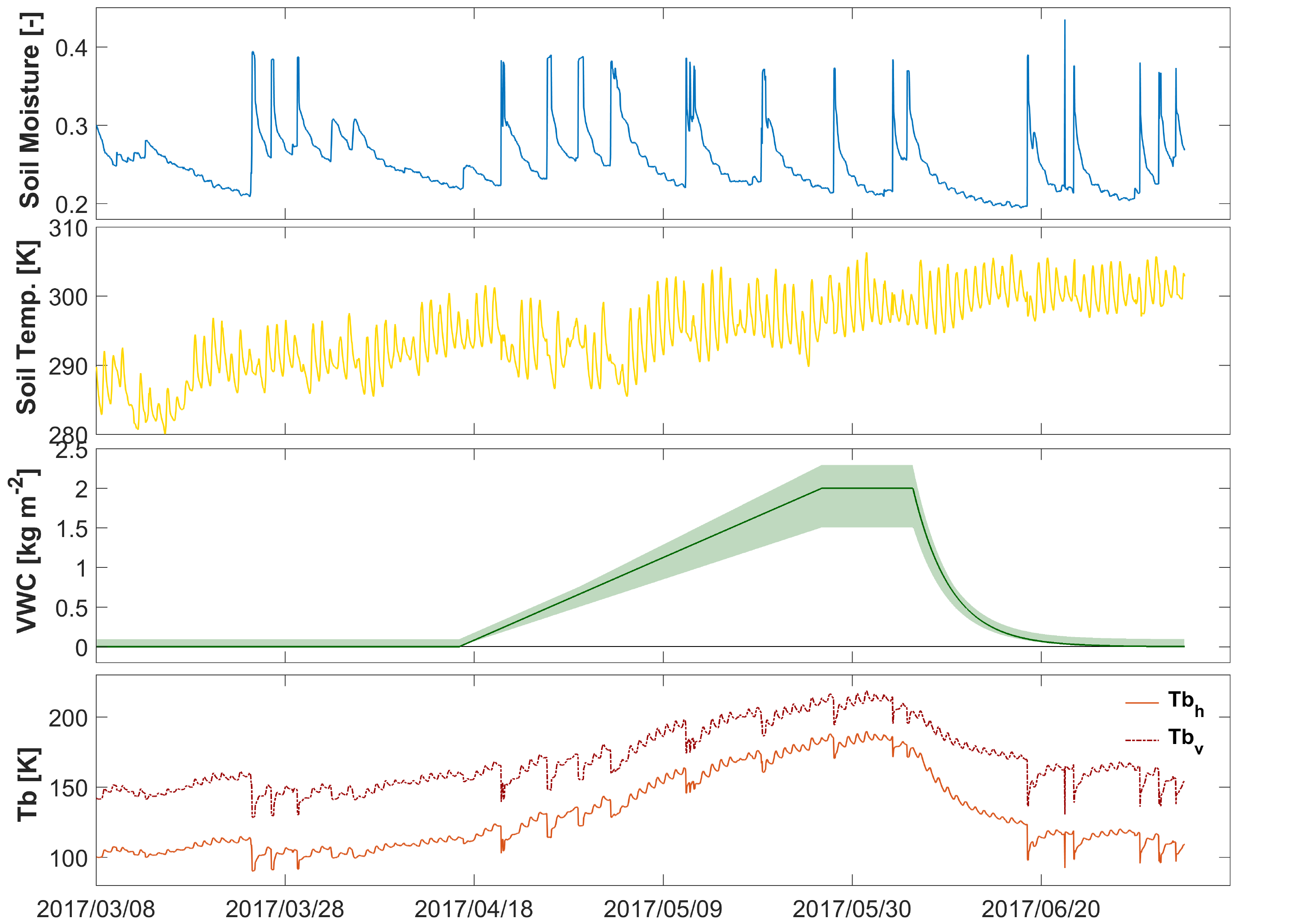}
\par\end{centering}
\caption{The hourly time series of surface soil moisture (first row), soil
temperature (second row), vegetation water content (third row), and
simulated brightness temperatures at 1.4 GHz for both horizontal (Tb\protect\textsubscript{h})
and vertical polarization (Tb\protect\textsubscript{v}) (fourth row). The data are from a SCAN gauge station in Arkansas, United States. The soil moisture and temperature are measured in depth 0.05 cm. The assumptions about the VWC and its uncertainties, shown in
shaded area, are synthetic. \label{fig:8}}
\end{figure}

To examine the results of the algorithm for a windowed retrieval, we use
surface soil moisture and temperature time series that are obtained
from a gauge station (N~35$^{\circ}$13$'$, W92$^{\circ}$55$'$) of the Soil Climate Analysis Network (SCAN) in Arkansas, United States (Figure \ref{fig:8},
first and second rows). We confine our consideration to 120 days of
hourly surface soil moisture and temperature observations at depth
5 cm from 03/08/2017 to 07/06/2017. In this period, the volumetric
soil moisture content changes from 0.194 to 0.435, which results in
$0.60\leq r_{Hr}\leq0.70$ and $0.44\leq r_{Vr}\leq0.57$ using the dielectric model \citep{Mironov_2009}. In the site, the first 20 cm of the soil is largely silty soil with clay content 4.1\%. 

To construct an inversion scenario, we hypothetically assumed that
the true VWC remains zero during the first 40 days and linearly increases
to 2 kg\,m\textsuperscript{-2} in the next 40 days. Then, the VWC
remains constant for 10 days and decays exponentially in the last 30 days with
a rate of 0.25 kg\,m\textsuperscript{-2}\,d\textsuperscript{-1}.
We consider 15\% multiplicative uncertainty around the VWC (i.e.,
VWC\textsubscript{min}=0.75~VWC, VWC\textsubscript{max}=1.15~VWC)
to define the box constraints for the vegetation transmissivity values.
When the VWC is zero, we consider a non-symmetric bound from zero
to 0.10 kg~m\textsuperscript{-2} (Figure \ref{fig:8}, third row).
Even though, the reconstructed problem might not be fully realistic
as the signals of soil moisture and VWC are not actually coupled, the
subsequent inversion experiment sheds light on how an a priori assumption
about smoothness in temporal dynamics of the VWC can be used for improved
soil moisture retrieval. 

For windowed retrievals, the standard form of the CMCA in equation \ref{eq:12} can be expanded as follows:
\begin{equation}
\begin{aligned}
& \underset{\mathbf{r}_{pr},\, \mathbf{\gamma}}{\text{minimize}}
& & \left\Vert \mathbf{e_{\tau}}-f(\mathbf{r}_{pr},\,\boldsymbol{\gamma})\right\Vert _{2}^{2}+\lambda_{1}\left\Vert \mathbf{r}_{pr}\right\Vert _{2}^{2}+\lambda_{2}\left\Vert \mathbf{D}\boldsymbol{\gamma}\right\Vert _{2}^{2} \\
& \text{subject to}
& & \boldsymbol{r}_{pl}\preceq\mathbf{r}_{pr}\preceq \boldsymbol{r}_{pu}\,\,,\,\,\boldsymbol{\gamma}_{l}\preceq\boldsymbol{\gamma}\preceq\boldsymbol{\gamma}_{u}.\label{eq:13}
\end{aligned}
\end{equation}
where $\mathbf{e}_{\tau}=\left(e_{1},\,\ldots,e_{t}\right)^{T}\in\mathbb{R}^{t}$,
$\mathbf{r}_{pr}=\left(r_{Hr1},\ldots,r_{Hrt},r_{Vr1},\ldots,r_{Vrt}\right)^{T}\in\mathbb{R}^{2t}$,
$\boldsymbol{\gamma}=\left(\gamma_{1},\,\ldots,\gamma_{t}\right)^{T}\in\mathbb{R}^{t}$,
and 
\begin{equation}
\mathbf{D}=\begin{bmatrix}1 & -2 & 1 & \cdots & 0\\
\vdots & \ddots & \ddots & \ddots & \vdots\\
0 & \cdots & 1 & -2 & 1
\end{bmatrix}\in\mathbb{R}^{(t-2)\times t}.
\end{equation}

Here, we chose a second order derivative operator for $\mathbf{D}$
to account for the smoothness in temporal variability of the VWC.
This choice enables to project a large
body of the time series of the VWC to zero and results in a smaller
2-norm than a first-order derivative. Choosing the first-order derivative
often results in a piece-wise linear retrievals between the soil moisture
jumps (not shown here). Note that the above
formulation allows for the use of two different regularization
parameters, which is necessary as the 2-norm of $\mathbf{r}_{rp}$ and $\mathbf{D}\gamma$ are significantly different and need to be scaled properly. Through a trial and error, we chose $\lambda_{1}$=1e-7 and $\lambda_{2}$=5e+2. To make the retrieval experiment computationally tractable, we solved problem \ref{eq:13} for non-overlapping windows of 10 days.

\begin{figure}[t]
\begin{centering}
\includegraphics[width=1.0\textwidth]{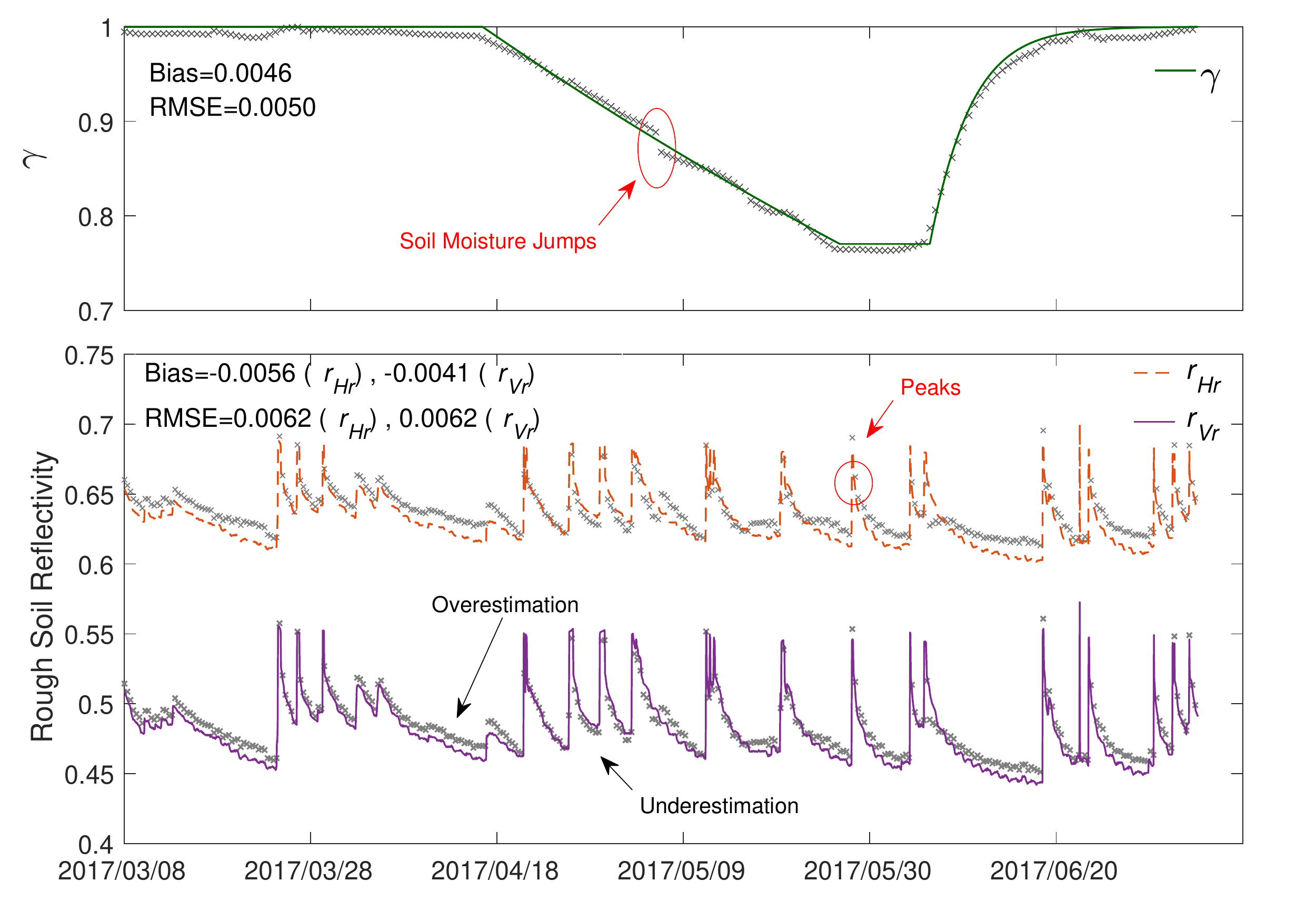}
\par\end{centering}

\caption{The CMCA windowed retrievals of the vegetation transmissivity ($\gamma$,
top row) and the rough surface soil reflectivity values ($r_{Hr}$
and $r_{Vr}$, bottom row) for the shown boundary conditions in Figure
\ref{fig:9}. Retrievals of $\gamma$ are sensitive to the jumps in
soil moisture content\textendash especially when the VWC is relatively high.
The results indicate that the peaks of the soil reflectivity values are
well retrieved, while the the reflectivity values are slightly overestimated
at the end of the soil moisture recession limbs. \label{fig:9}}
\end{figure}

The results show minor bias of less than 6\% in retrievals of reflectivity and transmissivity
values (Figure \ref{fig:9}). We observe that the peaks are captured well. However, when the soil moisture decaying limb is relatively long, the retrievals often overestimate the low values of soil moisture at the end of the limb, which is consistent with our previous finding
about the effects of the preferential solution space in LS inversion
of the $\tau$-$\omega$ model. As previously noted, we also observe
that the retrievals of soil reflectivity values in vertical polarization
are less biased. 

The relative RMSE in retrievals of transmissivity and reflectivity values is
around 3 and 6\% of their variability range. Overall, the experiment
reaffirms that the quality of the retrievals depends on accurate
characterization of the bounds and magnitude of the VWC. For example,
the RMSE for retrieval of $r_{Vr}$ reduces from 0.011 to 0.0034 (70\%),
when the mean of the VWC decreases from 2.5 to 0.1 kg\,m\textsuperscript{-2}.
When the multiplicative uncertainty factor is increased from 5 to
30\%, the RMSE of $r_{Vr}$ is also increased by $\sim70$\% from 0.0042
to 0.007. We need to note that, even though this relative increase
seems to be large, it is only about 3\% of the bound width of $r_{Vr}\in\left[0.44,\,0.57\right]$. The largest errors in retrieval of VWC often occur, when soil moisture suddenly jumps due to a precipitation event over vegetative surfaces (Figure \ref{fig:9}, first row). 

\subsection{Implementation for the SMAP Retrievals \label{subsec:Implementation-for-SMAP}}

In this subsection, we elaborate on implementation of the proposed algorithm for retrieval
of surface soil moisture and VWC using the SMAP observations. We confine our consideration to the SMAP observations over CONUS, where we can use multi-layer soil characteristics at a resolution of 1 km.
\begin{figure}[H]
\begin{centering}
\includegraphics[width=0.47\textwidth]{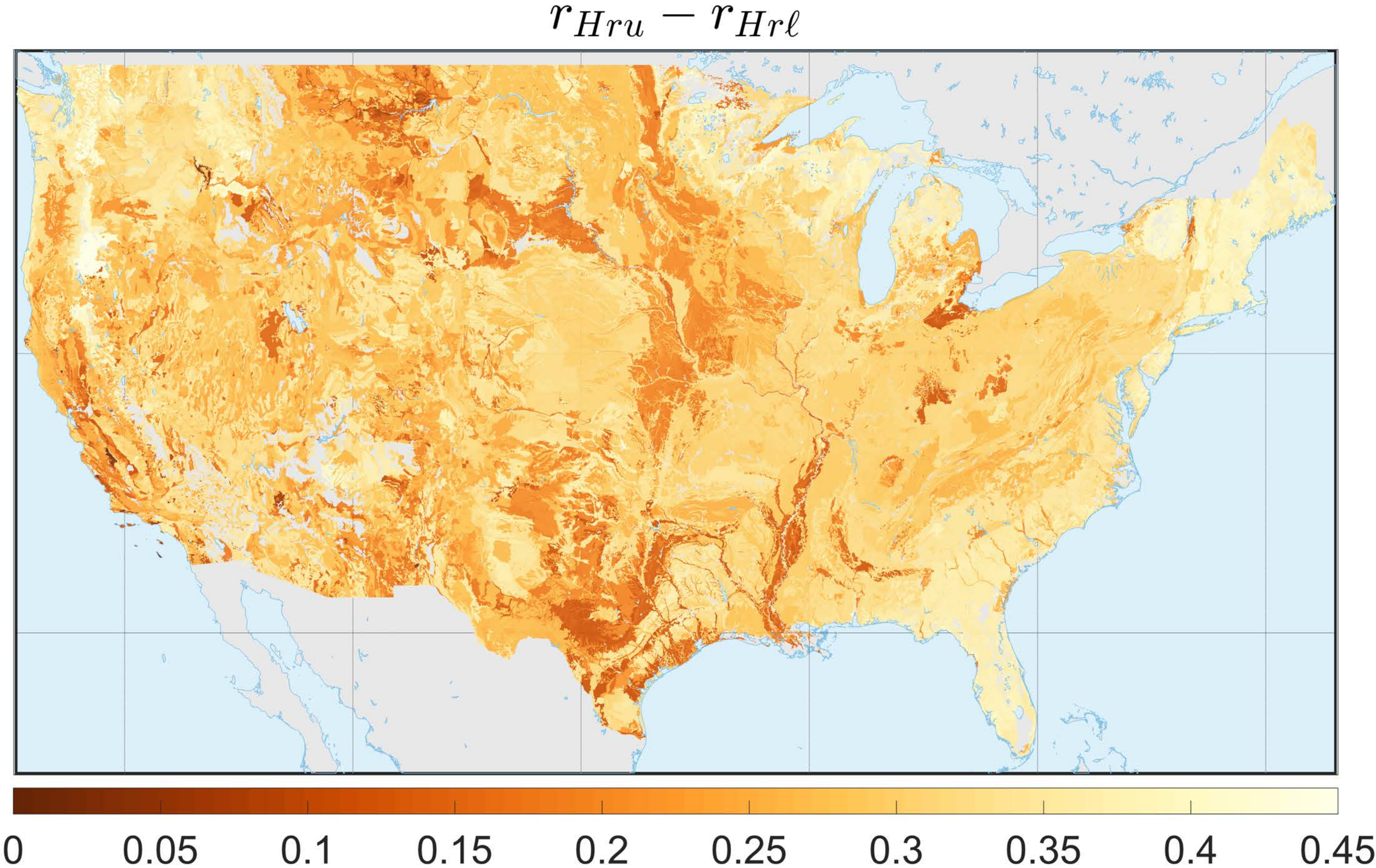}
\includegraphics[width=0.47\textwidth]{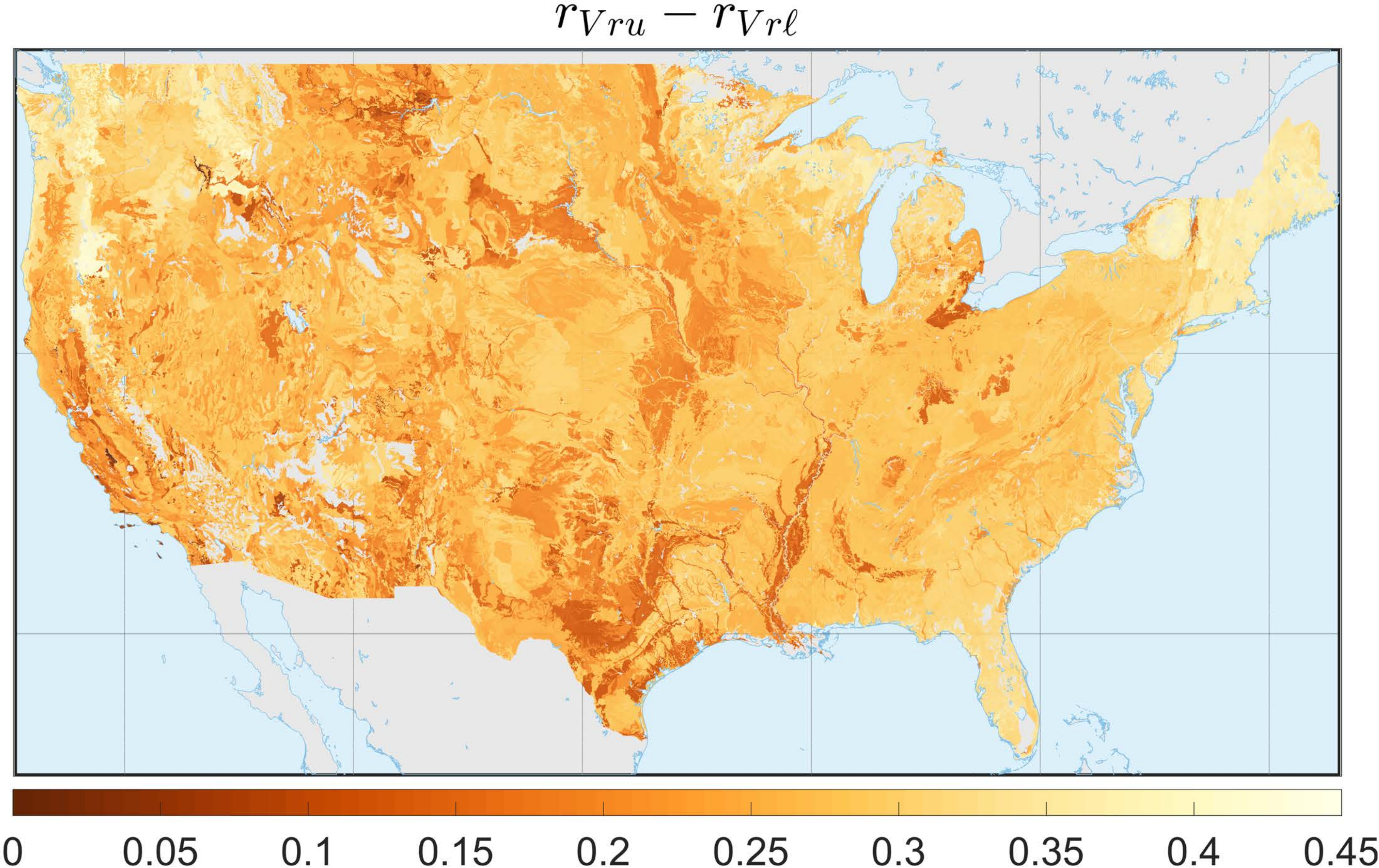} \\
\bigskip{}
\includegraphics[width=0.47\textwidth]{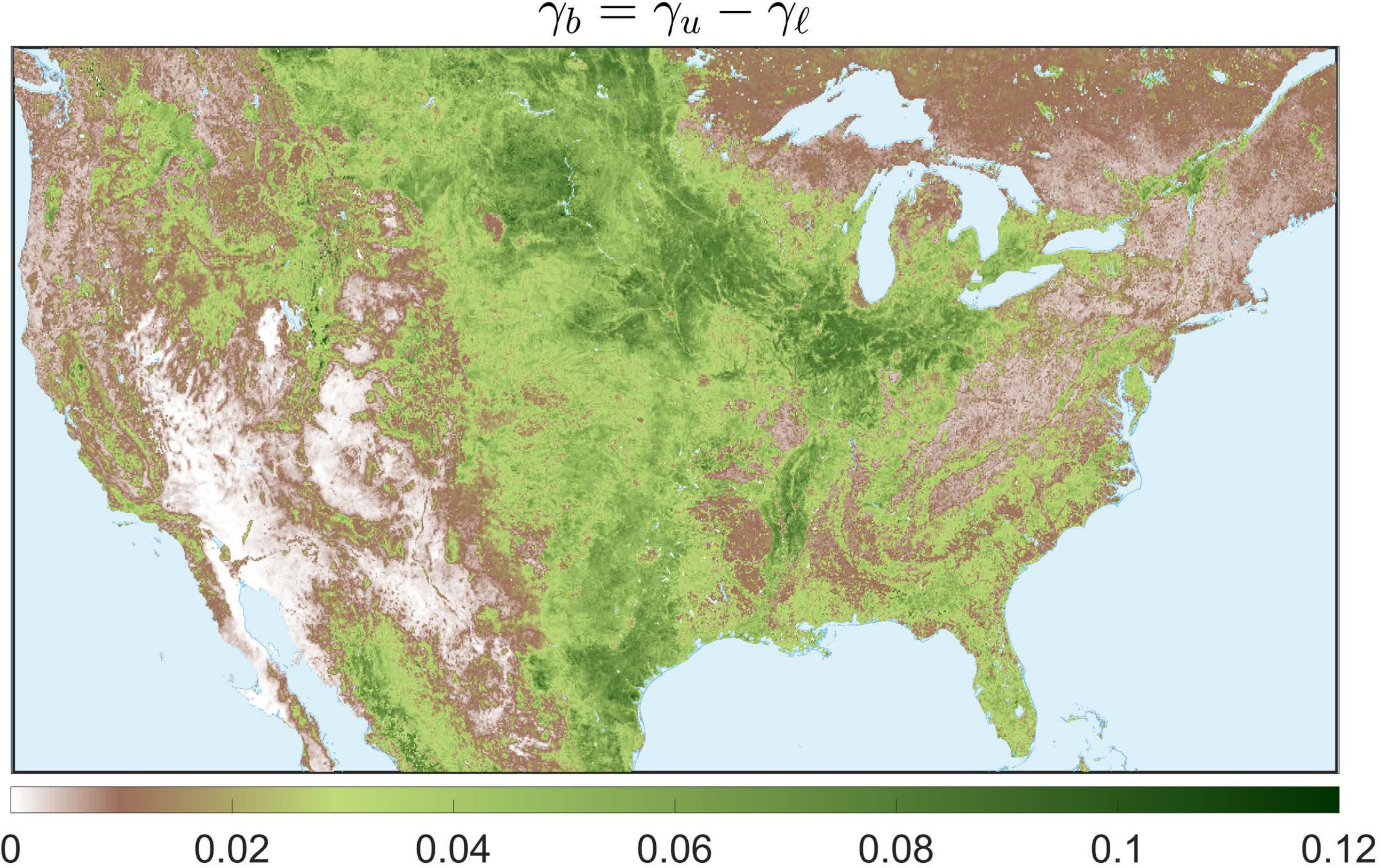}
\par\end{centering}
\caption{The static bounds of the rough surface soil reflectivity at $f=1.4$ GHz for horizontal ($r_{Hru}-r_{Hr\ell}$) and vertical ($r_{Vru}-r_{Vr\ell}$) polarization channels
over the CONUS at 1-km grid, where we assumed that the
soil moisture varies between the permanent wilting point and the soil
porosity. The monthly climatology of the vegetation transmissivity
bound ($\gamma_{b}=\gamma_{u}-\gamma_{\ell}$, second row) in month of June at resolution 0.05$^{\circ}$. \label{fig:10}}
\end{figure}
We use the soil dataset by \citet{Miller1998} that contains
compiled information from the State Soil Geographic (STATSGO) data
by the Natural Resources Conservation Service (NRCS) of the United
States Department of Agriculture. This dataset contains information
about soil texture classes, clay fraction, and porosity at grid resolution
1 km for 11 layers from surface to the depth of 2\,m. The first
layer represents the top soil from 0 to 5\,cm depth. To compute the
reflectivity bounds, we consider that the soil moisture varies
between the permanent wilting point and the soil porosity. Clearly,
this bound could be tightened in the future, for example
by assuming that the soil moisture varies between irreducible water
content and natural saturation.

\begin{wrapfigure}{o}{0.58\columnwidth}%
\begin{centering}
\includegraphics[width=0.58\textwidth]{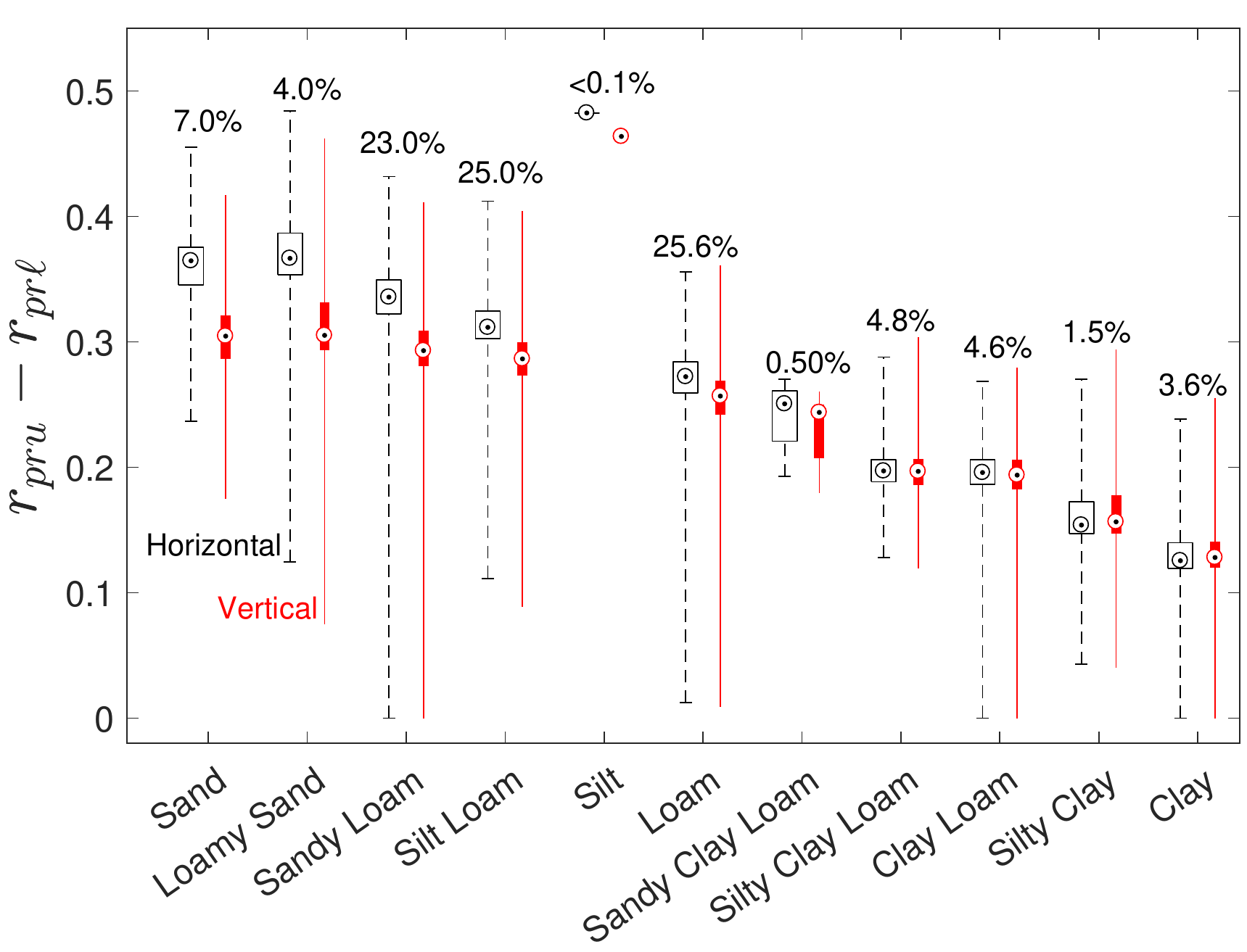}
\par\end{centering}
\caption{Reflectivity bounds for soil types over the CONUS at horizontal
(black dashed lines, $r_{Hru}-r_{Hr\ell}$) and vertical (solid red
lines, $r_{Vru}-r_{Vr\ell}$) polarization. The central
point is the median, the boxes span between the first and third quartiles,
and the whiskers are the min and max values.
The numerics are the areal percentage of the surface soil
types. \label{fig:11}}
\end{wrapfigure}

Figure \ref{fig:10} (first row) shows the static bounds of the vertically and horizontally polarized rough
surface soil reflectivity values at frequency 1.4 GHz. Over the CONUS, the
most abundant surface soil types are the loam (25.6\%), silt loam
(25.0\%) and sandy loam (23.0\%). The areal percentage of other surface
soil types remain below 6\%. The used soil data report almost zero
percentage of silt and sandy clay soils. Figure \ref{fig:11} shows
the box plot of the computed reflectivity bounds for the existing
soil types. As is evident, the median and width of the reflectivity bounds reduce when the clay content increases. It appears that there is not any significant difference between the width of the bounds for the horizontal and
vertical polarization. The widest bounds belong to the sandy loam,
loamy sand, and loam\textemdash largely due to the dynamic range of their
soil moisture and clay content. The distribution of the bounds is
often asymmetric with negative skewness.

Unlike the static nature of the soil reflectivity bounds, the bounds
on vegetation transmissivity can be defined dynamically. To that end,
we use 16-day NDVI data at resolution 0.05$^{\circ}$ from the Moderate
Resolution Imaging Spectroradiometer (MODIS) sensor on board the Terra
satellite \citep[MOD13C1-V6,][]{Didan_2015}. The pixel-level monthly
maximum and minimum values are computed using all available data from
2002 to 2017. The monthly timescale is chosen to address slow changes in VWC while providing sufficiently tight bounds for the
retrievals. We used the relationships by \citet{Jackson1999} and
\citet{Hunt1996} to convert the NDVI to foliage and stem water content
respectively. Then the VWC is transformed to the vegetation transmissivity
by assuming $b=0.1$, where $\tau=b\text{\,VWC}$, $\gamma=\exp\left(-\tau\,\sec\phi\right)$
and $\phi=40^{\circ}$. The bound on climatology of the vegetation
transmissivity ($\gamma_{max}-\gamma_{min}$) in the month of June is
shown in Figure \ref{fig:10} (bottom row). It is clear that the cultivated croplands
and natural grasslands show maximum amount of monthly variability
during a growing month, while the bound width is minimal for example over
the deciduous broadleaf forest of the Appalachian Mountains. 
\begin{figure}[H]
\begin{centering}
\includegraphics[height=5.38cm]{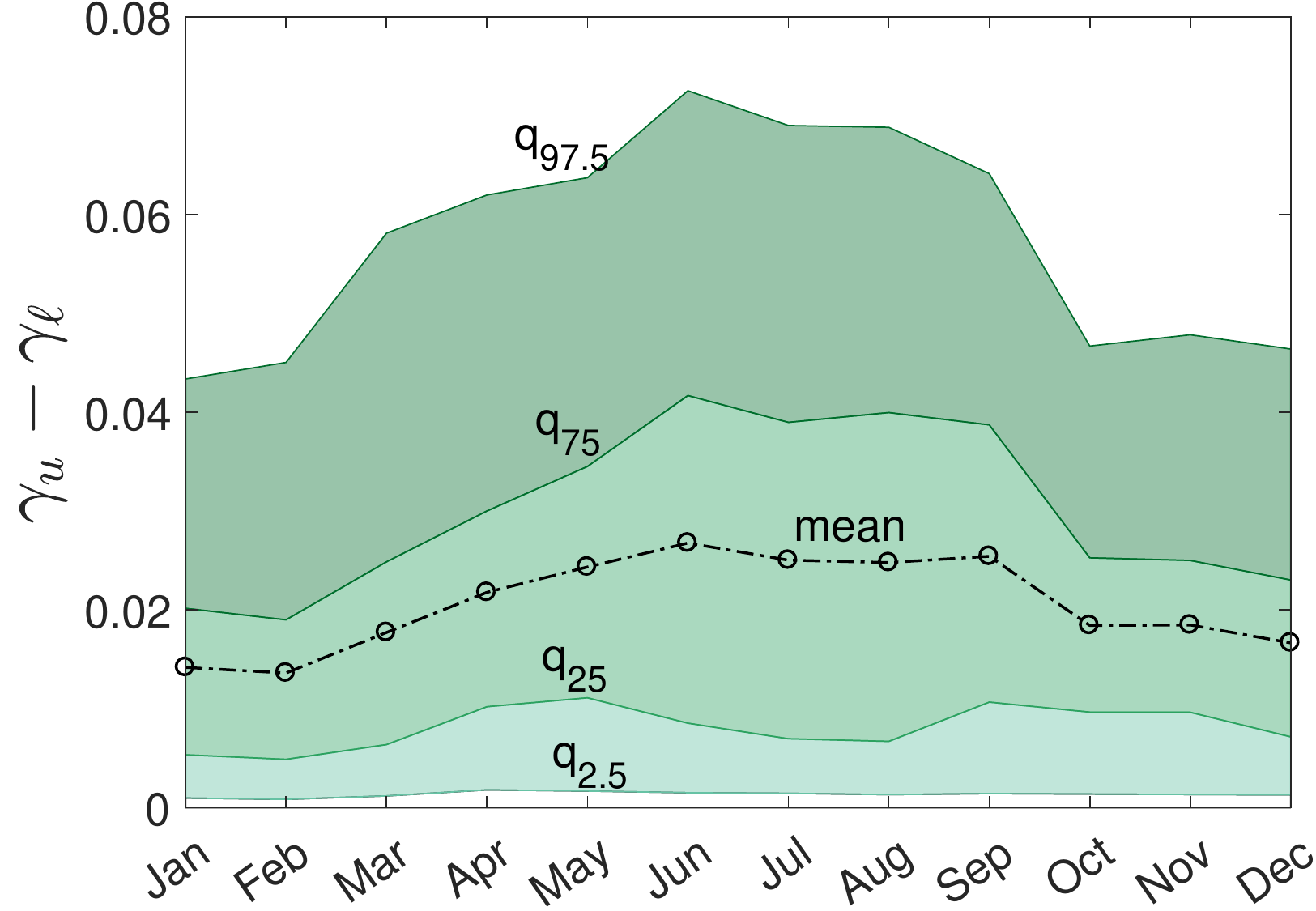}$\quad$\includegraphics[height=5.3cm]{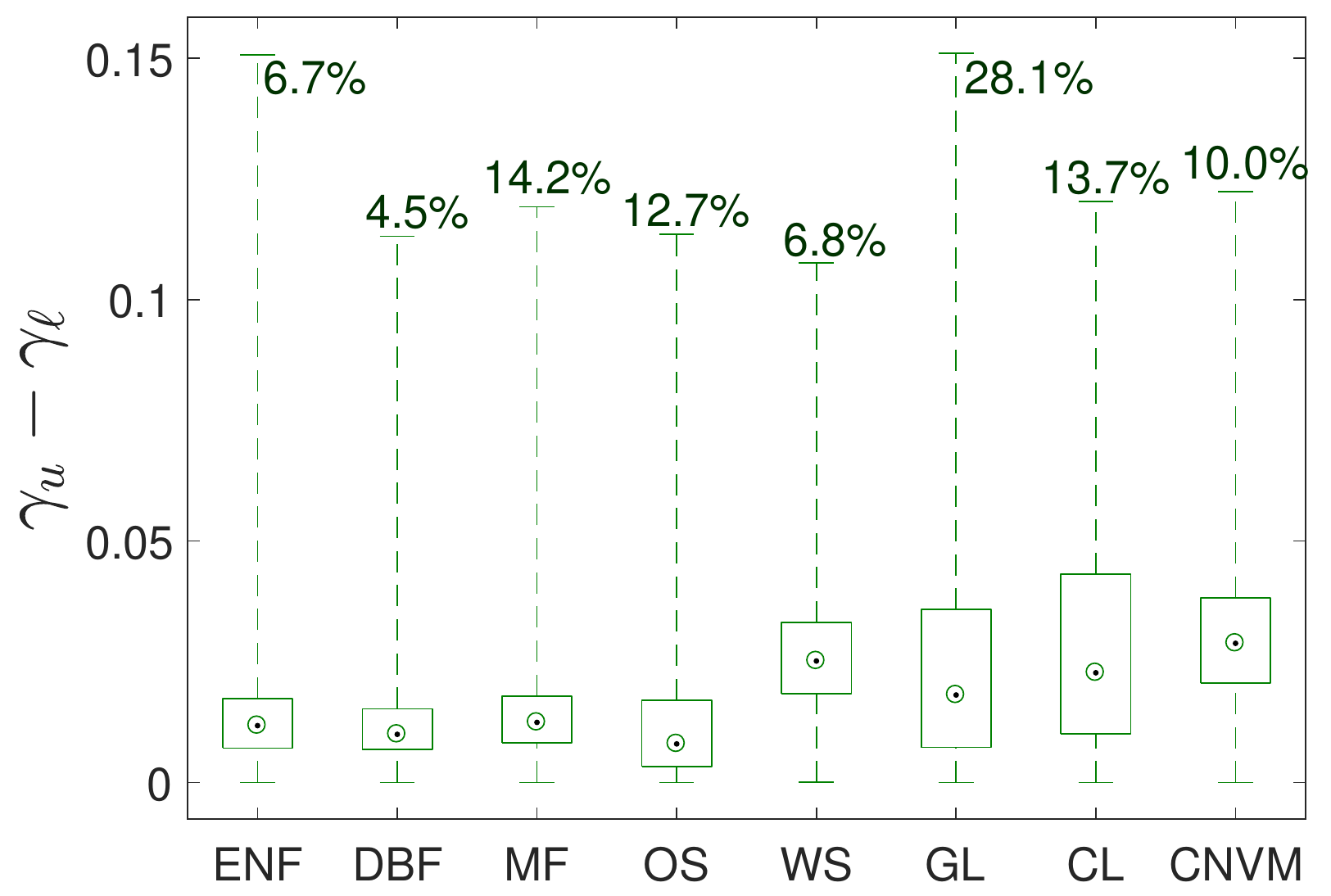}
\par\end{centering}
\caption{climatology of the vegetation transmissivity bounds
($\gamma_{b}=\gamma_{u}-\gamma_{\ell}$) for different months of the
year (left) and dominant IGBP land cover types (right) over the CONUS.
The land cover types include the evergreen needleleaf forest (ENF),
deciduous broadleaf forest (DBF), mixed forest (MF), open shrublands
(OS), woody savannas (WS), grasslands (GL), croplands (CL), and cropland/natural
vegetation mosaic (CNVM), where the numerical values denote the areal percentage of each land cover type. \label{fig:12}}
\end{figure}
For improved understanding of the spatiotemporal dynamics of the bounds
of the vegetation transmissivity, Figure \ref{fig:12} (left) shows
monthly changes in the spatial mean of \mbox{$\gamma_{b}=\gamma_{u}-\gamma_{l}$}
over the CONUS.  The bounds are relatively tight as the width is smaller than 0.10.
We can see that the difference between percentile 97.5 and 2.5 increases
from $\sim0.04$ to 0.08 from dormant to growing months and reaches
to its maximum around June and July. 

We also study the distribution of $\gamma_{b}$ over dominant land cover
types (Figure \ref{fig:12}, right). To that end, the classification
by the International Geosphere-Biosphere Programme (IGBP) is adopted.
This classification over the CONUS is obtained from the MODIS combined
product MCD12C1 provided by the NASA's Land Processes Distributed
Active Archive Center. The maximum values of $\gamma_{b}$ reaches to
0.15 over the evergreen needless forests and grasslands. The grasslands and croplands cover more than 40\% of the CONUS and have the widest interquartile range of $\gamma_{b}$; even though, the extremum values are not significantly different than the other surface types. 
\begin{figure}[t]
\begin{centering}
\includegraphics[width=0.70\textwidth]{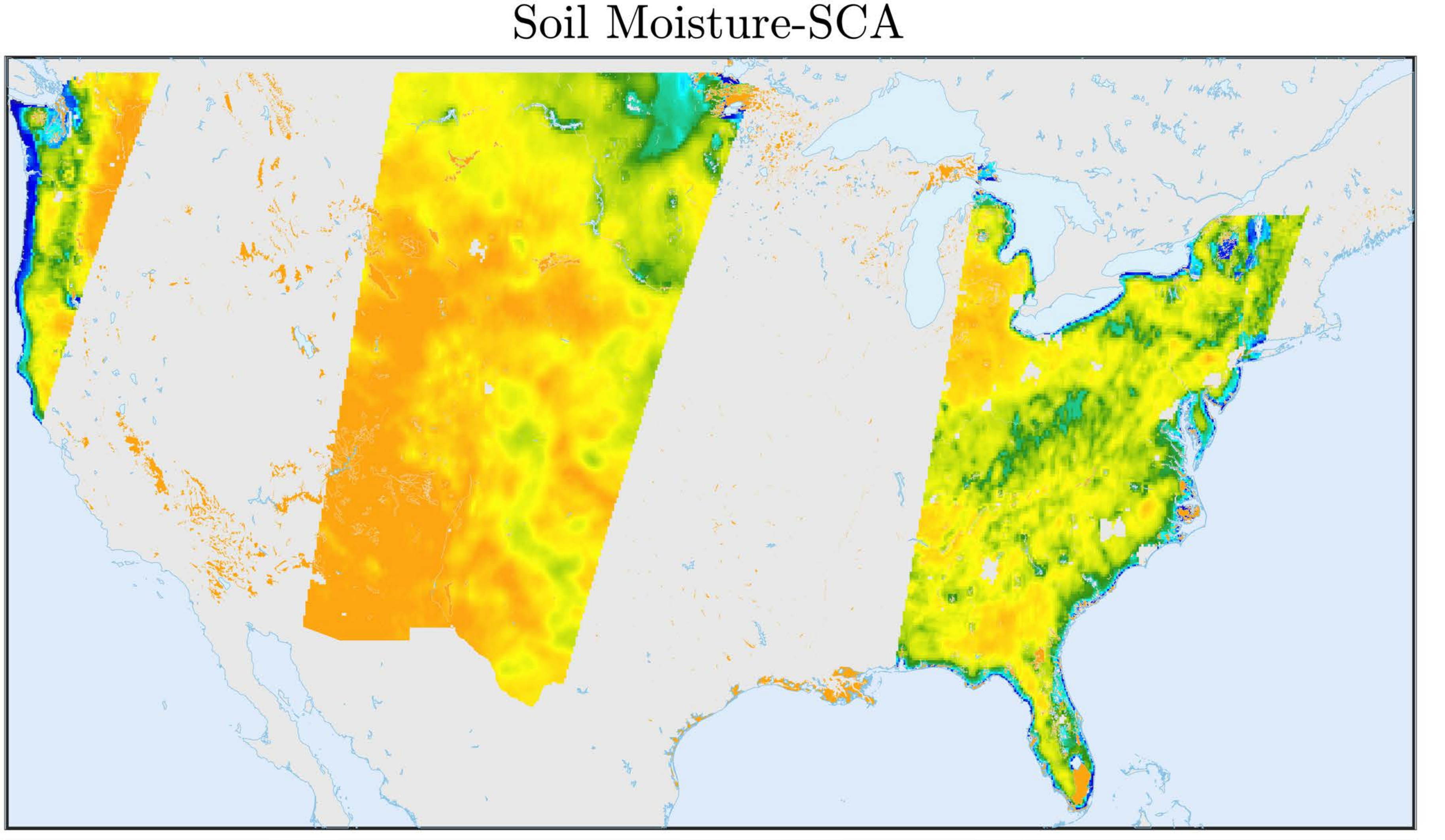} \\
\includegraphics[width=0.70\textwidth]{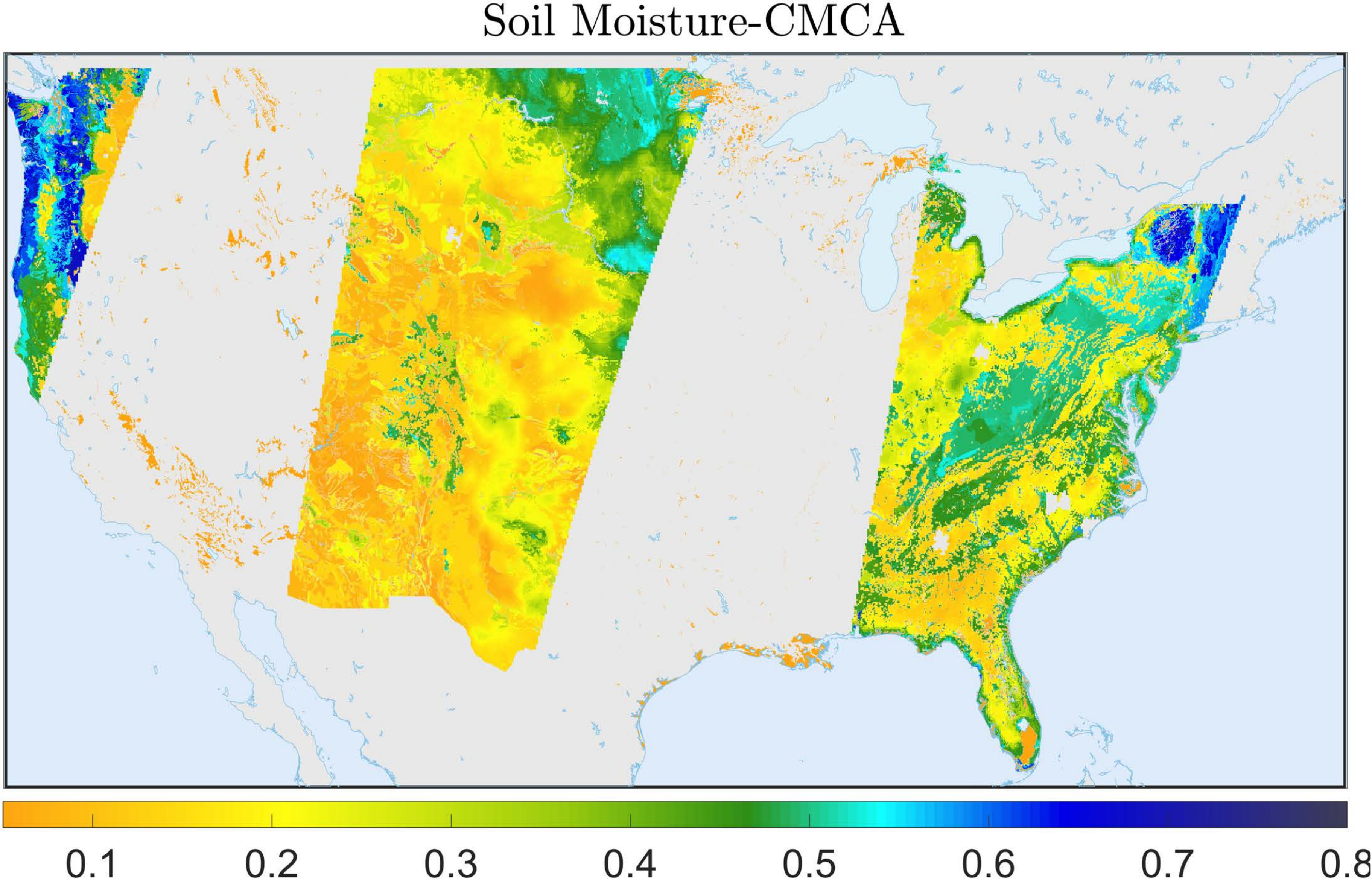}
\par\end{centering}
\caption{The enhanced SMAP (first row) and CMCA (second row) soil moisture retrievals on 06/01/2016, at nominal resolution 9 and 1\,km, respectively. \label{fig:13}}
\end{figure}

We examined the initial results of the algorithm for a retrieval experiment
using SMAP observations over the CONUS. We focused on the enhanced
level-III SMAP radiometric data and soil moisture retrievals \citep{ONeill2018} with nominal
grid resolution of 9\,km on 06/01/2016 (Figure \ref{fig:13}, first
row), which are derived by interpolating the values from the nearest
6 instantaneous fields of view (IFOV) of the SMAP radiometer footprints. As previously explained, the VWC content in the SMAP data (Figure \ref{fig:15}, first row) is a 10-day climatology from MODIS data that is averaged to match the soil moisture resolution at 9 km. One advantage of the CMCA approach is  that the resolution of the retrievals is not only a function of the native resolution of radiometer but also the resolution
of constraints. For the case of the SMAP retrieval over the CONUS,
we have static constraints for the soil reflectivity values at resolution
1\,km and dynamic constraints of vegetation transmissivity at resolution 1 to
5\,km. Here, we map the SMAP radiometric observations and vegetation transmissivity bounds onto a 1\,km grid using the nearest neighbor interpolation to conduct retrievals at this nominal resolution.

\begin{figure}[t]
\begin{centering}
\includegraphics[width=0.70\textwidth]{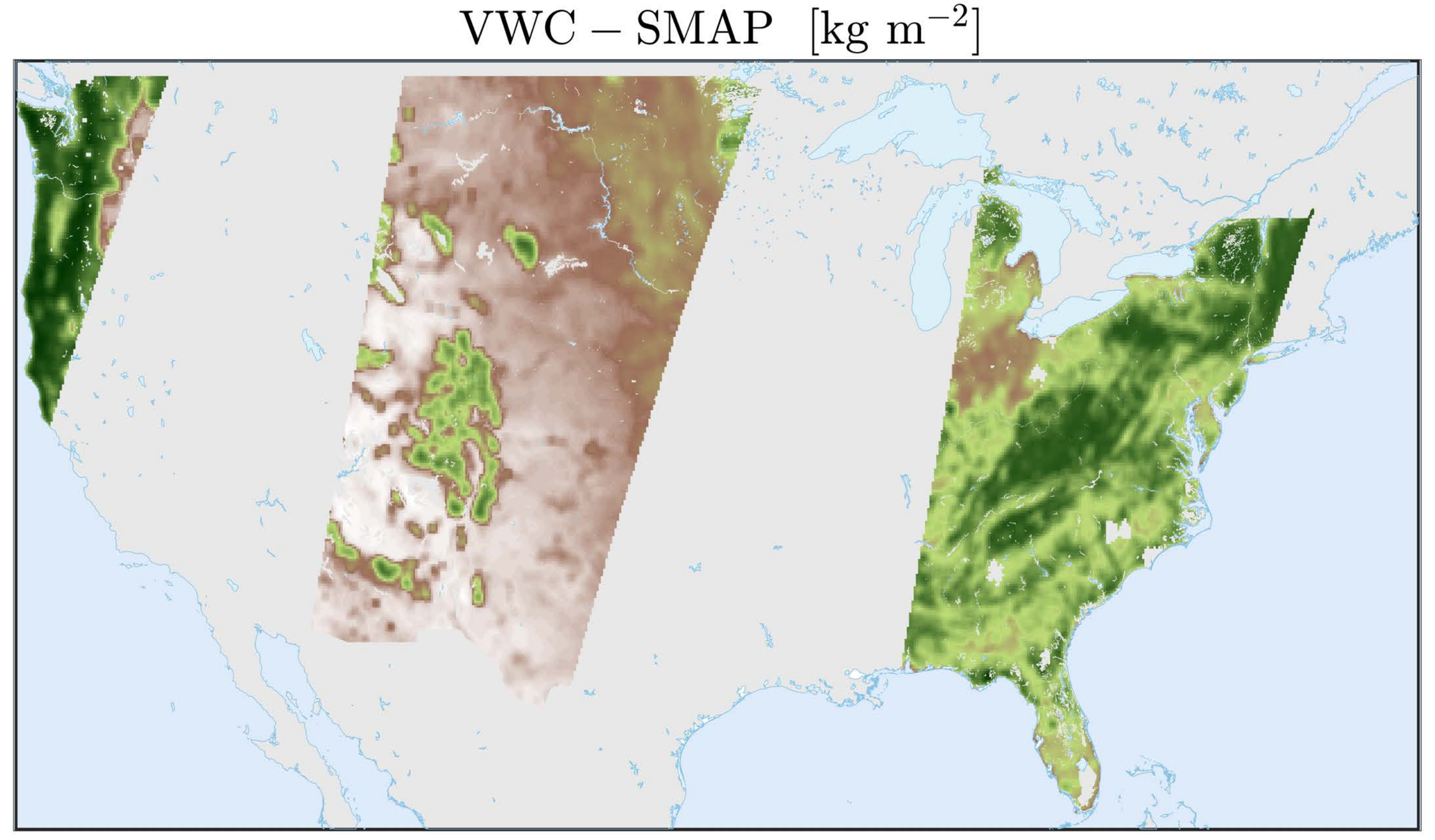} \\
\includegraphics[width=0.70\textwidth]{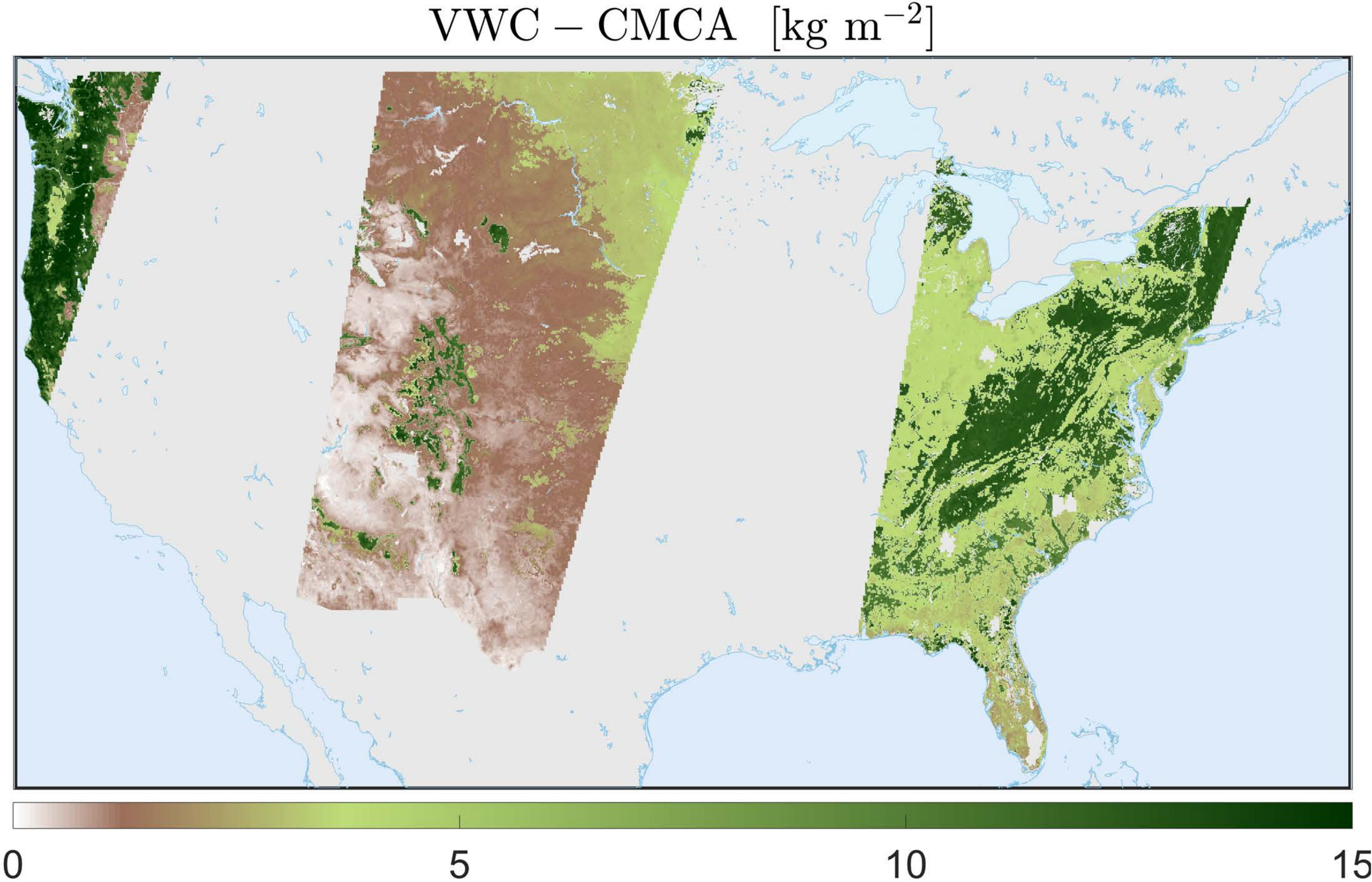}
\par\end{centering}
\caption{The ancillary VWC data (first row)  at resolution of 9\,km reported in enhanced SMAP products and the results of CMCA retrievals of the VWC (second
row) at 1\,km resolution on 06/01/2016. \label{fig:14}}

\end{figure}

The CMCA retrievals of soil moisture and VWC retrievals at 1\,km grid resolution are shown in Figure \ref{fig:13} and \ref{fig:14} (second rows). The CMCA retrievals provide a high-resolution representation of both soil moisture and VWC, due to the high-resolution constraints. To quantify the retrieved extra high-resolution soil moisture details, Figure \ref{fig:15} shows the local variability of the retrievals at different scales, ranging from 2 to 64 km. The retrieved soil moisture fields are basically broken down into non-overlapping neighborhoods of size 2 to 64 km and then the expected values of the standard deviation are computed within those neighborhoods.

\begin{wrapfigure}{r}{0.55\columnwidth}%
\begin{centering}
\includegraphics[width=0.55\textwidth]{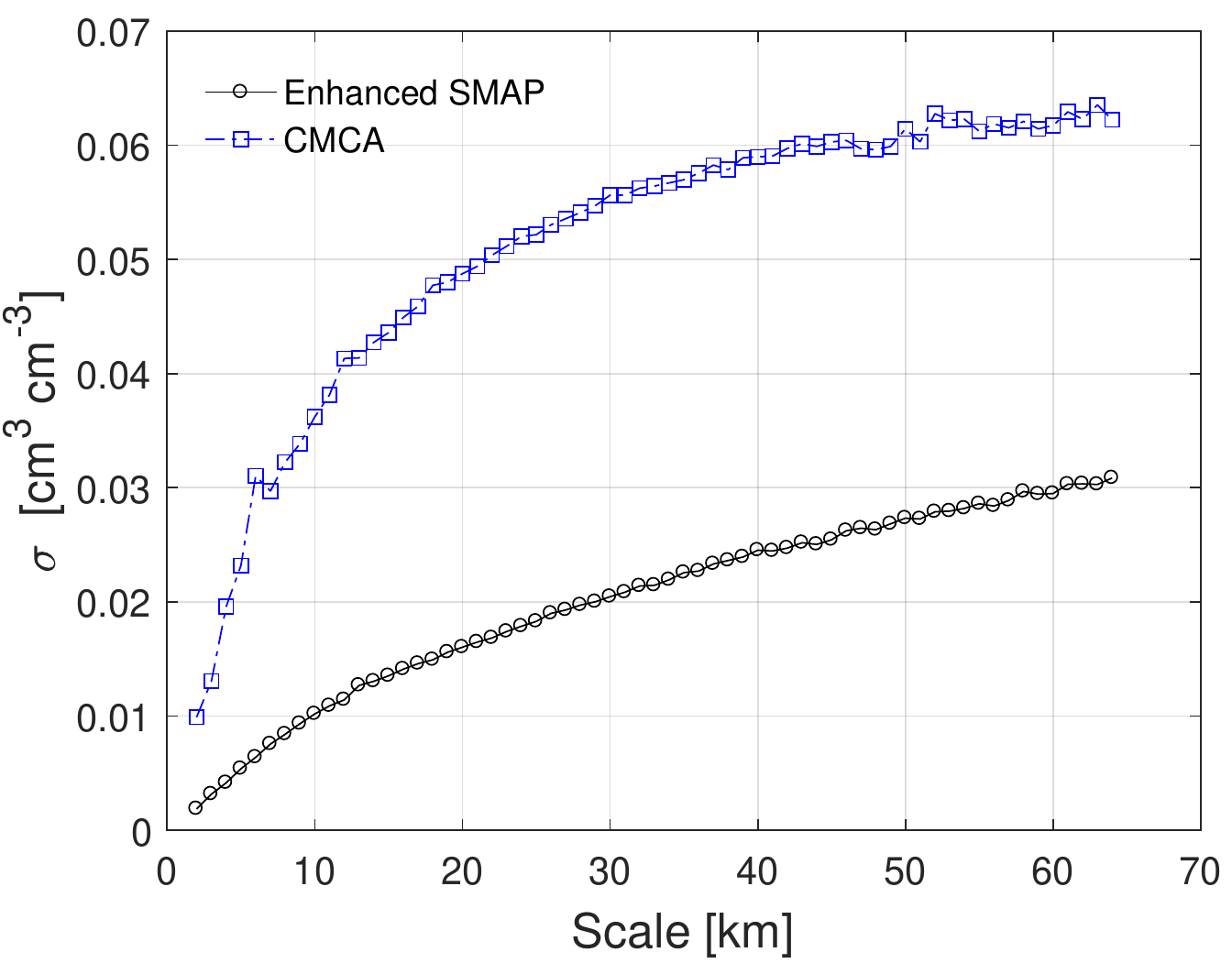}
\par\end{centering}
\caption{Standard deviations ($\sigma$) of the soil moisture local fluctuations in Figure \ref{fig:14} at different scales. \label{fig:15}}

\end{wrapfigure}%
The computed standard deviations are measures of the variability of the soil moisture fluctuations around its local mean at different scales. As expected, the standard deviation increases as a function of scales and reaches to a
plateau, when the analysis scale becomes greater than the largest dominant
mode of the soil moisture spatial variability. In the presented
retrieval experiment, the calculated standard deviations 
of the CMCA retrievals are 2 to 3 times larger than their counterparts for the enhanced product. The difference increases from 0.01 to 0.03 [cm\textsuperscript{3} cm\textsuperscript{-3}] as the analysis scale increases. This difference is an indication that some extra high-resolution details can be recovered by the CMCA. However, understanding the signal to noise ratio of this high-resolution details requires a thorough comparison of the retrievals against ground-based observations. 

An important observation is that the CMCA slightly overestimates the SMAP product. This overestimation is pronounced over the evergreen needleleaf forests of the Pacific coast ranges, Colorado forests, and the deciduous broadleaf forests of the Appalachians, where the VWC is generally above 5 kg\,m\textsuperscript{-2} and thus the $\tau$-$\omega$ model cannot fully explain the radiometric signature of the soil moisture. However, there are also overestimations in croplands of the Northern Minnesota and North Dakota, where the VWC is less than 5 kg\,m\textsuperscript{-2}.

\subsection{SMAP Validation Experiments} \label{sec:validation}

\begin{wrapfigure}{o}{0.60\columnwidth}%
\begin{centering}
\includegraphics[width=0.60\textwidth]{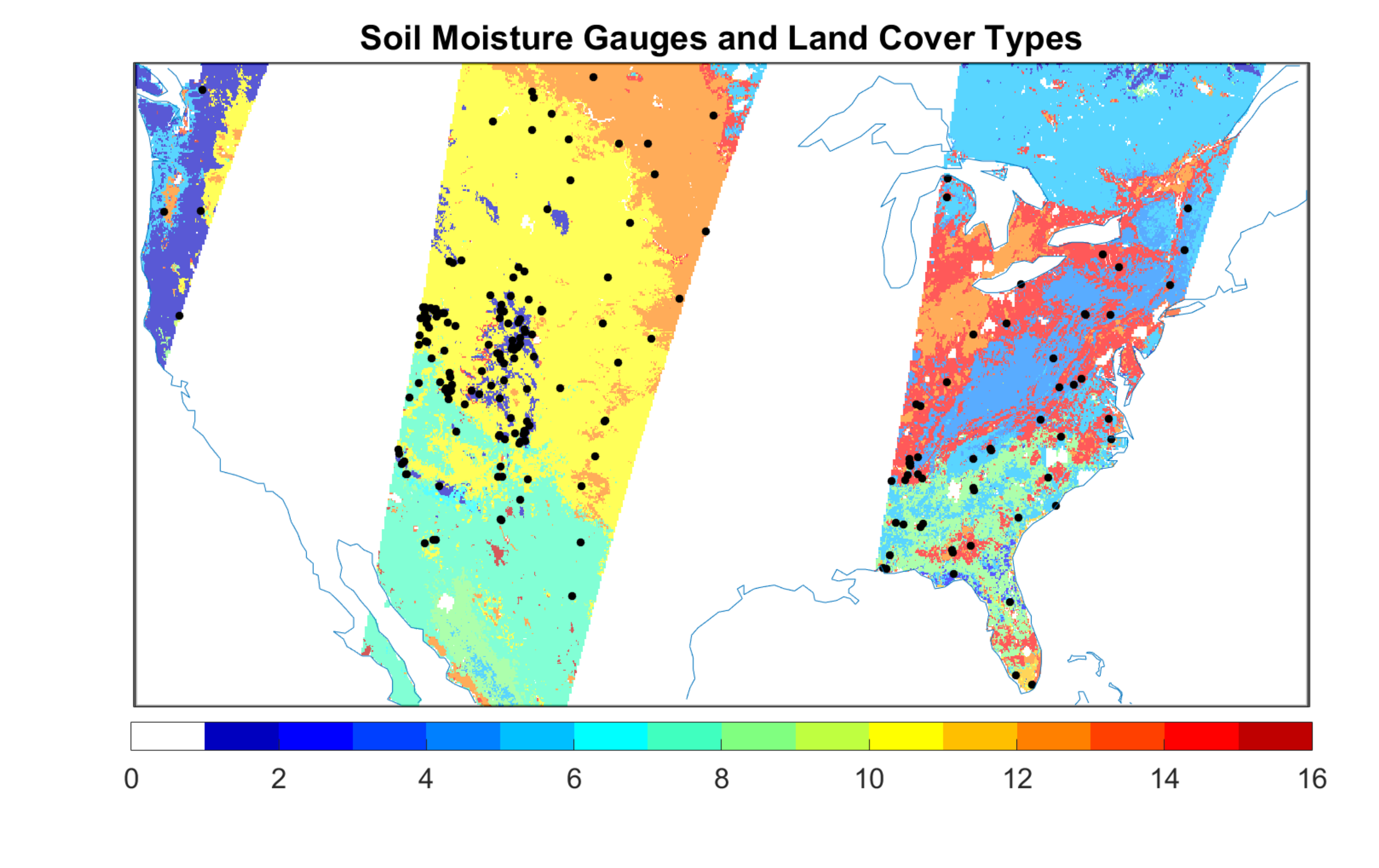}
\par\end{centering}
\caption{Location of the soil moisture gauge stations from the ISMN database and the IGPB land cover types, where labels 10, 12 and 14 denote grasslands, croplands, and cropland/natural vegetation mosaic, respectively.  \label{fig:16}}
\end{wrapfigure}

To validate the retrieved soil moisture, we used the soil moisture gauge data from the International Soil Moisture Network (ISMN) database \citep{Dorigo2011,Dorigo2013}. Screening that data, we identified 206 gauges with good quality flags that provide soil moisture observations within the SMAP swath width on 06/01/2016 (Figure \ref{fig:16}). These gauges are from U.S. Climate Reference Network (USCRN), Soil Climate Analysis Network (SCAN), Snow Telemetry (SNOWTEL), Interactive Roaring Fork Observation Network (iRON), and Cosmic-ray Soil Moisture Observing System (COSMOS). The gauge observations, reported at Coordinated Universal Time, are interpolated linearly onto the SMAP scanning times. 

Among these gauges, 110 gauges are over grasslands and croplands where the VWC is less than 5 kg\,m$^{-2}$. Comparing these gauge data with the nearest pixels of soil moisture retrievals in Figure \ref{fig:13} indicates that the bias in CMCA is around +0.02, while it is around -0.025 in the SMAP retrievals and the standard deviation of both retrievals is around 0.14 (Figure \ref{fig:16}).

\begin{figure}[t]
\begin{centering}
\includegraphics[width=1\textwidth]{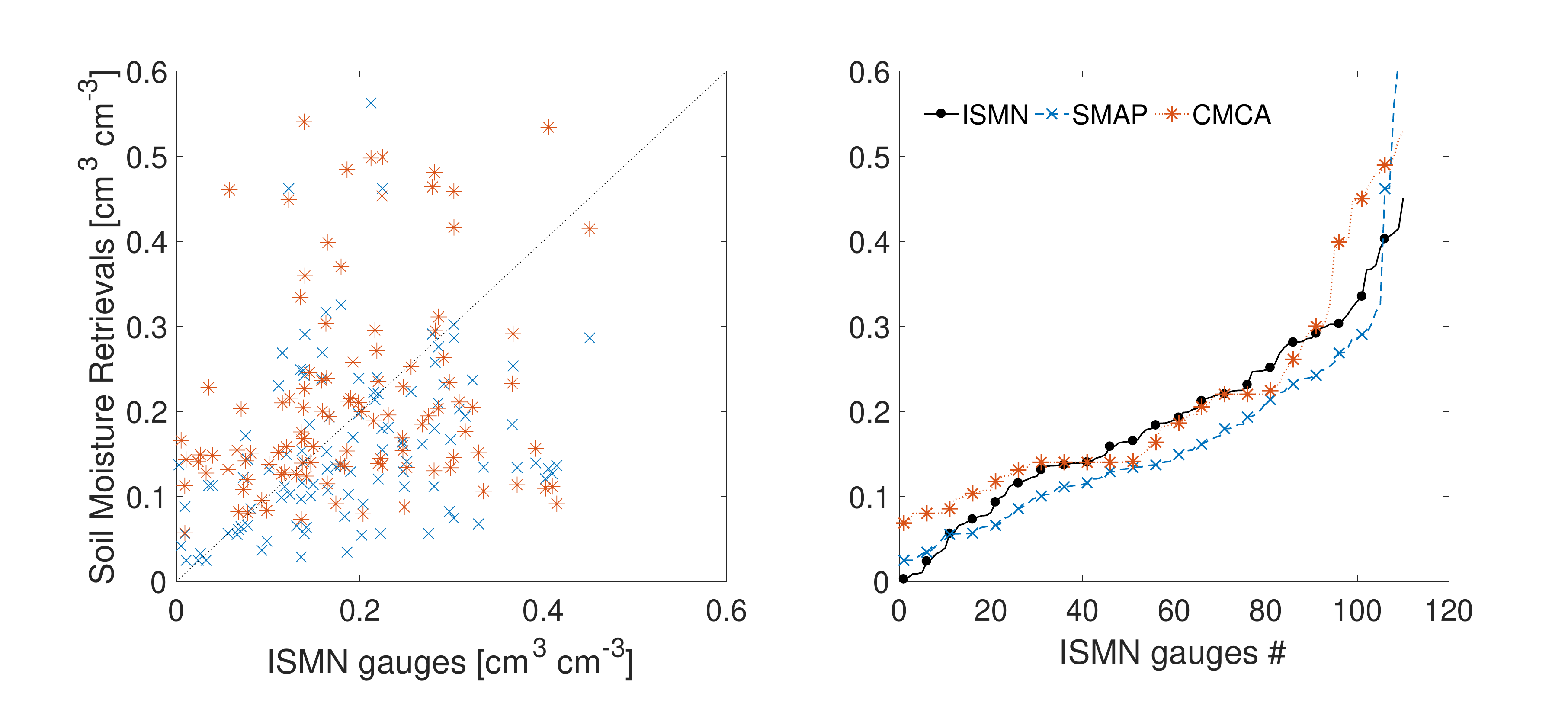}
\par\end{centering}
\caption{ISMN soil moisture gauge data versus CMCA and SMAP retrievals over grass and croplands (left) and sorted data (right) showing the pattern of over and underestimations. \label{fig:17}}
\end{figure}%

The results show that CMCA overestimates (underestimate) the soil moisture when the gauge measurements are below 0.15 (above 0.3). The overestimation could be due to the fact that the lower bound of the soil moisture retrievals is set to the soil permanent wilting point, which is 0.05 and 0.30 [cm$^3$ cm$^{-3}$] in sandy and clayey soils (Table \ref{tab:1}). We also confined the upper bound of soil moisture retrievals to the soil porosity, which is generally higher than the soil natural saturation. Table \ref{tab:2} reports the sensitivity of the error statistics to the lower and upper bounds of soil moisture retrievals, which are shifted by two multiplicative parameters. 

\begin{table} \tiny
\resizebox{\textwidth}{!}{\begin{tabular}{|c|c|c|c|c|c|c|c|c|c|c|c|c|c|c|c|c|}
\hline 
$c_l$-$c_h$ & 1-0.80 & 1-0.90 & 0-0.7 & 0-0.80 & 0-0.90 & 0.7-0.7 & 0.80-0.80 & 0.85-0.85 & 0.90-0.90\tabularnewline
\hline 
\hline 
Bias & 0.008 & 0.014  & -0.029 & -0.022 & -0.015 & -0.016 & -0.004 & 0.002 & 0.007\tabularnewline
\hline 
RMSE & 0.126 & 0.135  & 0.118 & 0.127 & 0.137 & 0.117 & 0.125 & 0.130 & 0.135\tabularnewline
\hline 
\end{tabular}}

\caption{Error statistics of the CMCA retrievals against the gauge data for different bounds, where the soil permanent wilting point and porosity are multiplied by $c_l$ and $c_h$ respectively. \label{tab:2}}

\end{table}

\begin{wrapfigure}{o}{0.58\columnwidth}%
\begin{centering}
\includegraphics[width=0.58\textwidth]{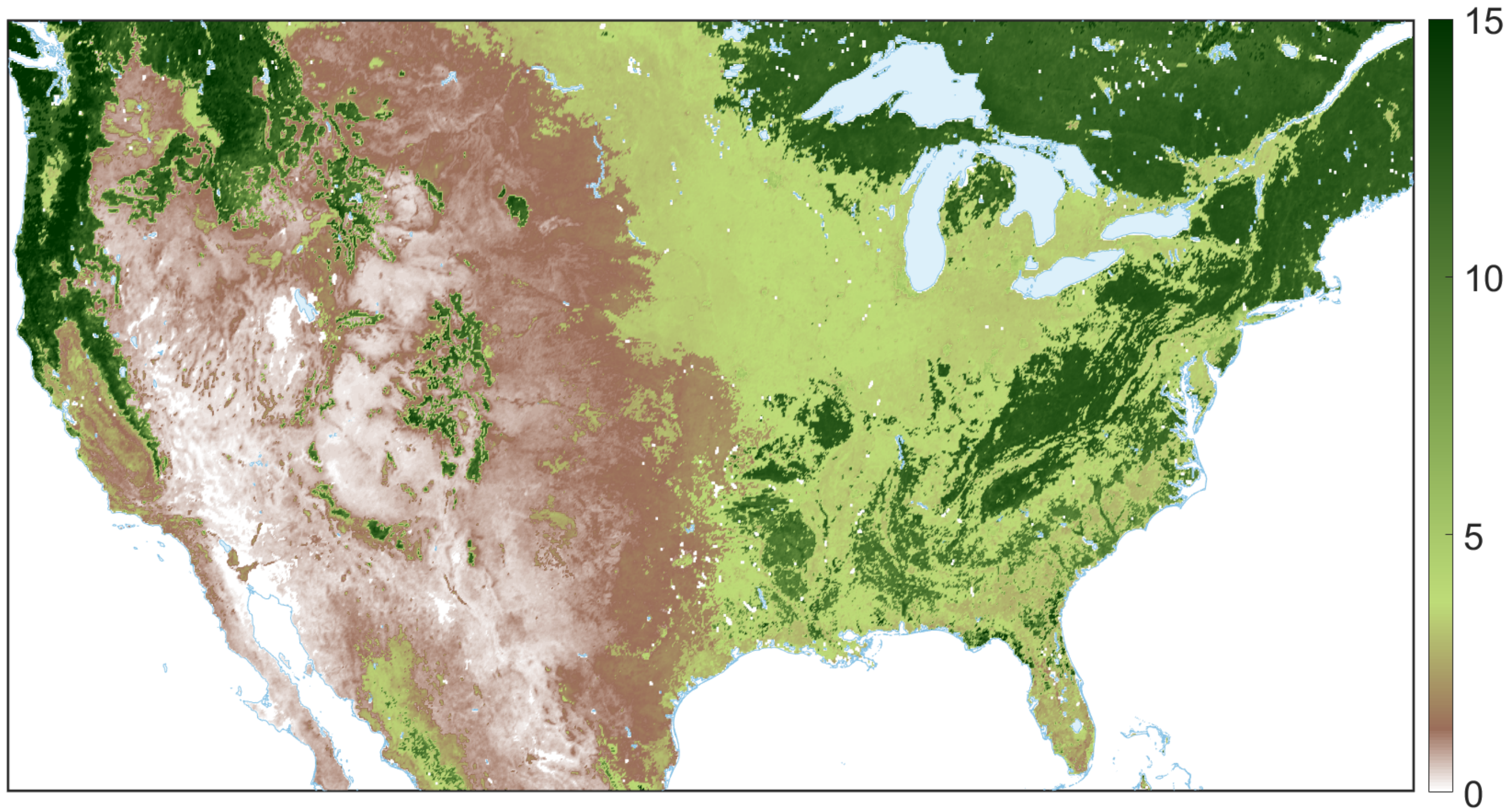}
\par\end{centering}
\caption{The derived VWC from 16-day MODIS NDVI data on 06/01/2016. \label{fig:18}}
\end{wrapfigure}%

To validate the retrievals of the VWC we use as a reference the derived VWC from the 16-day NDVI data available on 05/24 and 06/09/2016 (Figure \ref{fig:18}). To this end, we first obtained a temporal weighted average of the NDVI values and then derived the VWC at  $0.05^{\circ}$-degree.  We see that there are differences between the observations and the climatology of the VWC (Figure \ref{fig:14}, first row). The climatology map underestimates the observed VWC, especially over grass and croplands, which leads to overestimation of the vegetation transmissivity and thus underestimation of soil moisture. This issue could be exacerbated due to coarse-graining of the NDVI data, knowing that the VWC is quadratically related to the NDVI \cite{Jackson1999}. Visual inspection shows that the retrieved vegetation transmissivity values by CMCA are consistent with the MODIS observations and capture well the VWC of croplands. The bias and RMSE values for the CMCA retrievals are 0.38 and 1.5 kg\,m$^{-2}$ over the pixels where the VWC is below 5 kg\,m$^{-2}$. These statistics for the climatology of VWC are -0.83 and 1.8 kg\,m$^{-2}$ for the bias and RMSE.

\section{Discussion and Concluding Remarks\label{sec:Discussion-and-Concluding}}

The existing single channel algorithms (SCA) for passive microwave
soil moisture retrievals often rely on climatology of VWC as a priori knowledge, 
which is obtained through ancillary NDVI data. On the other hand, the double channel algorithms (DCA) retrieve both soil moisture and VWC simultaneously without using any
a priori knowledge about the soil type and vegetation density, which could lead to biased results.

We presented a new constrained passive microwave soil moisture retrieval
algorithm that closes the gaps between these two widely used algorithmic
approaches. This algorithm conditions its retrievals to the static physical properties
of surface soil and can account for uncertainties in climatology of
the VWC. A smoothing norm regularization is proposed to extend the
algorithm for soil moisture retrievals over a window of time to
formally account for slow varying dynamics of VWC in grass and croplands. In this paper, we validated
the outputs of the algorithm through a series of synthetic and an initial ground-validation experiments.
Using the SMAP observations over the CONUS, we demonstrated that the algorithm
can lead to super-resolved retrievals of soil moisture and VWC given
high-resolution constraints. 

Any operational application of CMCA requires a thorough calibration of the parameters based on validation against ground-based gauge observations. From an algorithmic
stand point, the quality of the CMCA retrievals depends highly on
the a priori bounds that characterize feasible range of the surface soil reflectivity
and transmissivity of the overlying vegetation. Study of the climatology of the ground-based
soil moisture data and use of much higher-resolution NDVI data (e.g.,
250 m from the MODIS sensor) could lead to improved characterization
of the constraints. Since we observed that the calculated bounds
are not distributed uniformly, probabilistic consideration of the bounds could also lead to improved retrievals. Moreover, the accuracy of the inversion of the $\tau$-$\omega$ model
is related multiplicatively to the accuracy of the surface soil temperature.
Therefore, extension of the approach to a combined retrieval
algorithm that optimally fuses multiple sources of reanalysis surface
temperatures could be another line for future research. 

\section*{Acknowledgments}

The authors acknowledge the support (NNX16AM12G) from the NASA's Science
Utilization of the Soil Moisture Active-Passive (SMAP) Mission through
Dr. J. Entin. The enhanced SMAP soil moisture data (version 2) are provided courtesy
of the NASA Distributed Active Archive Center (DAAC) at National Snow
and Ice Data Center (NSIDC, \url{https://nsidc.org/data/smap/smap-data.html}).
The MODIS data are from the Goddard Earth Sciences and Information
Service Center (\url{https://disc.sci.gsfc.nasa.gov/mdisc}/) and
the Land Processes Distributed Active Archive Center by the USGS (\url{https://lpdaac.usgs.gov/data_access/data_pool}). The authors also thank Dara Entekhabi at Massachusetts Institute of Technology for providing the codes for the used dielectric model.  

\section*{References}





\bibliographystyle{model1-num-names}




\end{document}